\documentclass[12pt]{article}
\usepackage{epsf,feynmp}
\def\href#1#2{#2}
\newcommand{\ezero}{\setcounter{equation}{0}}
\def\theequation{\arabic{section}.\arabic{equation}}
\begin{document}
\begin{fmffile}{wwfm2}\unitlength=1mm
\thispagestyle{empty}
\onecolumn
\begin{flushright}
\end{flushright}

\begin{flushleft} 
{\tt
DESY 98-047
\\
hep-ph/9805355
\\
May 1998}
\end{flushleft}

\vfill

\begin{center}{\huge 
Off-shell $W$ pair production
\vspace*{.5cm}
\\
with anomalous couplings:
\vspace*{.5cm}
\\
 The {\tt CC11} process
\vspace*{3.4cm}
     } 

\vfill

\noindent
{\Large {\it J.~Biebel}\/ and {\it T.~Riemann}}\\
\end{center}
\vspace*{1.0cm}
\begin{center}
\large
DESY Zeuthen
\\  Platanenallee 6, D-15738 Zeuthen, Germany
\end{center}

\vfill

\centerline{\large Abstract}

\bigskip

\normalsize
\noindent
The differential cross-sections for processes of the type
$e^+e^-\to(W^+W^-)\to l\nu q\bar{q}$ are determined with account of background 
contributions and of anomalous triple gauge boson coup\-lings.
Analytic expressions for
$\mbox{d}\sigma/\mbox{d}s_1\mbox{d}s_2\mbox{d}\cos\theta$,
where~$\theta$ is the production angle of the~$W$ boson,
are numerically integrated with the Fortran package {\tt GENTLE}. 
QED corrections are taken into account in the leading logarithmic
approximation.
The importance of the various contributions is studied 
for center-of-mass energies of 190~GeV, 500~GeV, and 1~TeV.

\vfill

\newpage

\section{Introduction}
\label{introduction}
\ezero

Since the establishment of the electroweak standard model
\cite{Glashow:1961ez}-
\nocite{Weinberg:1967pk}
\cite{Salam:1968rm} many precision
tests confirmed its validity in various respects.
One of the poorly investigated features is the non-Abelian nature of
gauge couplings.
$W$ pair production,
\begin{equation}
  e^+e^-\to W^+W^-,
\end{equation}
provides an excellent way to investigate the triple gauge boson self couplings.

\begin{figure}[b]
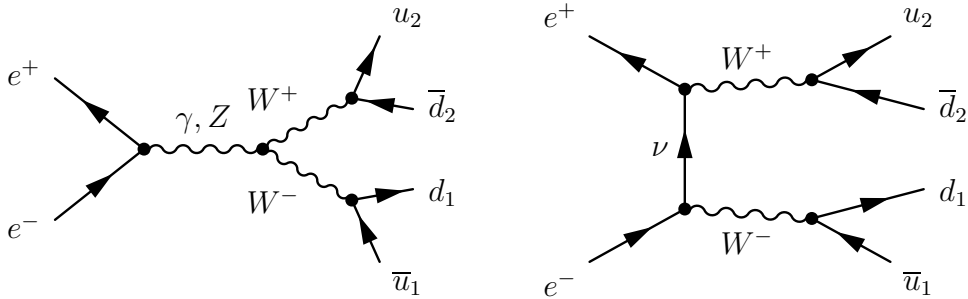

\begin{center}
  \begin{fmfchar*}(50,30)
    \fmfleft{a1,k1,a2,a3,k2,a4}
    \fmfright{p2,p1,p4,p3}
    \fmf{fermion,label=$$,l.s=left}{k1,v1}
    \fmf{fermion,label=$$,l.s=left}{v1,k2}
    \fmf{photon,label=$\noexpand\gamma,,Z$,tension=1.5,l.s=left}{v1,v2}
    \fmf{photon,label=$W^-$,l.s=right}{v2,v3}
    \fmf{photon,label=$W^+$,l.s=left}{v2,v4}
    \fmf{fermion}{p4,v4,p3}
    \fmf{fermion}{p2,v3,p1}
    \fmflabel{$e^-$}{k1}
    \fmflabel{$e^+$}{k2}
    \fmflabel{$d_1$}{p1}
    \fmflabel{$\noexpand\overline{u}_1$}{p2}
    \fmflabel{$u_2$}{p3}
    \fmflabel{$\noexpand\overline{d}_2$}{p4}
    \fmfdot{v1,v2,v3,v4}
  \end{fmfchar*}
\qquad\qquad
  \begin{fmfchar*}(50,30)
    \fmfleft{k1,k2}
    \fmfright{p2,p1,p4,p3}
    \fmf{fermion}{k1,v1}
    \fmf{fermion,label=$\noexpand\nu$,l.s=left}{v1,v2}
    \fmf{fermion}{v2,k2}
    \fmf{phantom}{k1,v1}
    \fmf{phantom}{k2,v2}
    \fmf{photon,label=$W^-$,tension=1.5,l.s=right}{v1,v3}
    \fmf{photon,label=$W^+$,tension=1.5,l.s=left}{v2,v4}
    \fmf{fermion}{p4,v4,p3}
    \fmf{fermion}{p2,v3,p1}
    \fmflabel{$e^-$}{k1}
    \fmflabel{$e^+$}{k2}
    \fmflabel{$d_1$}{p1}
    \fmflabel{$\noexpand\overline{u}_1$}{p2}
    \fmflabel{$u_2$}{p3}
    \fmflabel{$\noexpand\overline{d}_2$}{p4}
    \fmfdot{v1,v2,v3,v4}
  \end{fmfchar*}
\end{center}
\caption{\it The doubly resonating {\tt CC03} contributions to off-shell
  $W$ pair production.
\label{cc03diagram}
}
\end{figure}

First calculations of the cross-section for on-shell $W$ pair
production in a renormalizable theory, the standard model, were done
in the seventies \cite{Flambaum:1975wp,Alles:1977qv}.
Already before the formulation of the standard model the $W$ width was
estimated to yield sizeable effects if the $W$ boson is much heavier 
than the proton \cite{Tsai:1965hq}.
Due to the finite width, $W$ bosons decay immediately and the
production of four fermions is observed.
Production of off-shell $W$ pairs,
\begin{equation}
  e^+e^-\longrightarrow W^+W^- \longrightarrow \bar{f_2}f_2'f_1\bar{f}'_1,
\label{cc03p}
\end{equation}
through the three diagrams of figure \ref{cc03diagram} was calculated first in
\cite{Muta:1986is}.
Feynman diagrams without an intermediate $W$ pair will also contribute to the 
four fermion final states in~(\ref{cc03p}):
\begin{equation}
  e^+e^- \longrightarrow  f_1\bar{f}'_1\bar{f_2}f_2'.
\label{cc11p}
\end{equation}
They constitute the so-called irreducible background and are
experimentally not distinguishable from the signal diagrams.
Classifications and first studies may be found in 
\cite{Berends:1994pv,Bardin:1994sc}, and an overview in  
\cite{Beenakker:1996kt}. 
Further, photonic, electroweak, and QCD radiative corrections must be regarded
in order to achieve sufficient accuracy of numerical predictions.
A huge literature exists on this subject.
See e.g.~\cite{Bardin:1986fi}-%
\nocite{Jegerlehner:1986vs,Bohm:1988ck,Fleischer:1989kj,Denner:1990tx,%
Beenakker:1991sf,Dittmaier:1992np,Fleischer:1993nw,%
Beenakker:1997bp,Beenakker:1997ir}
\cite{Denner:1997ia}, and references therein.

The properties of triple boson vertices are investigated with different
approaches. 
Polarization amplitudes for the most general form of the $\gamma W^+W^-$
and $ZW^+W^-$ vertices compatible with Lorentz invariance were determined
in~\cite{Gaemers:1979hg,Hagiwara:1987vm}. 
Since then, many studies appeared on $W$ pair production with anomalous
couplings, see e.g. 
\cite{Hagiwara:1993ck}-%
\nocite{
Bilchak:1984ur,Jegerlehner:1994zp}
\cite{HarunarRashid:1994mp}.
For a recent overview, see e.g.~\cite{Gounaris:1996rz}.

\bigskip

The study of the physics of $W$ bosons is one of the main goals of LEP~2
and a future high-energy linear collider.
LEP~2 operates above the $W$ pair production threshold at about 161 GeV.
Several thousands of $W$ pairs will be produced and 
precise measurements of mass, width, and couplings of the $W$ boson
will become possible.
Later, at a future linear collider with an energy of 500~GeV or more
at high luminosity the number of produced $W$ pairs will be
even larger than at LEP~2. 
For a review see \cite{Accomando:1997wt}.

\begin{figure}
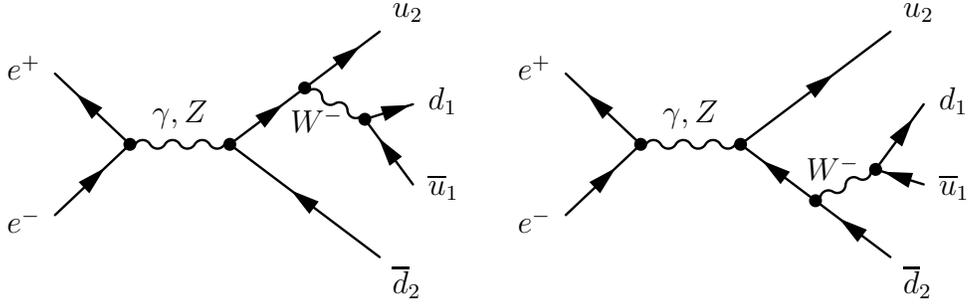

\begin{center}
  \begin{fmfchar*}(50,30)
    \fmfleft{a1,k1,a2,a3,k2,a4}
    \fmfright{p1,p2,p3,p4}
    \fmf{fermion,tension=2,label=$$,l.s=left}{k1,v1}
    \fmf{fermion,tension=2,label=$$,l.s=left}{v1,k2}
    \fmf{photon,label=$\noexpand\gamma,,Z$,tension=3,l.s=left}{v1,v2}
    \fmf{phantom}{p1,v2,p4}
    \fmffreeze
    \fmf{phantom}{v2,v3,p4}
    \fmffreeze
    \fmf{photon,label=$W^-\!\!$,tension=.8,l.s=right,l.d=2}{v3,v4}
    \fmf{fermion}{p1,v2}
    \fmf{fermion,tension=1.2}{v2,v3}
    \fmf{fermion}{v3,p4}
    \fmf{fermion,tension=.5}{p2,v4,p3}
    \fmflabel{$e^-$}{k1}
    \fmflabel{$e^+$}{k2}
    \fmflabel{$u_2$}{p4}
    \fmflabel{$\noexpand\overline{u}_1$}{p2}
    \fmflabel{$d_1$}{p3}
    \fmflabel{$\noexpand\overline{d}_2$}{p1}
    \fmfdot{v1,v2,v3,v4}
  \end{fmfchar*}
\qquad\qquad
  \begin{fmfchar*}(50,30)
    \fmfleft{a1,k1,a2,a3,k2,a4}
    \fmfright{p1,p2,p3,p4}
    \fmf{fermion,tension=2,label=$$,l.s=left}{k1,v1}
    \fmf{fermion,tension=2,label=$$,l.s=left}{v1,k2}
    \fmf{photon,label=$\noexpand\gamma,,Z$,tension=3,l.s=left}{v1,v2}
    \fmf{phantom}{p1,v2,p4}
    \fmffreeze
    \fmf{phantom}{p1,v3,v2}
    \fmffreeze
    \fmf{photon,label=$W^-\!\!\!$,tension=.8,l.s=left,l.d=2}{v3,v4}
    \fmf{fermion}{p1,v3}
    \fmf{fermion,tension=1.2}{v3,v2}
    \fmf{fermion}{v2,p4}
    \fmf{fermion,tension=.5}{p2,v4,p3}
    \fmflabel{$e^-$}{k1}
    \fmflabel{$e^+$}{k2}
    \fmflabel{$\noexpand\overline{d}_2$}{p1}
    \fmflabel{$\noexpand\overline{u}_1$}{p2}
    \fmflabel{$d_1$}{p3}
    \fmflabel{$u_2$}{p4}
    \fmfdot{v1,v2,v3,v4}
  \end{fmfchar*}
\end{center}
\caption{\it Four of the eight singly resonant contributions to off-shell $W$ pair
  production: the $d_2$-diagrams and the $u_2$-diagrams.
\label{cc11diagram}
}
\end{figure}

As mentioned, the process (\ref{cc11p}) may be classified by the final
state fermions.
In this article, we will treat the {\tt CC11} class, defined by two
requirements on the final state fermions: 
(i) they have to belong to two different weak isospin doublets
and  (ii) no electrons nor electron neutrinos are produced.
Besides the doubly resonating diagrams of figure~\ref{cc03diagram} there
are up to eight background diagrams of the types shown in
figure~\ref{cc11diagram}.
This depends on the number of neutrinos in the final state:
 $l_1\bar{\nu}_1\bar{l}_2\nu_2$,
$l_1\bar{\nu}_1\bar{q}q'$, $q_1\bar{q}'_1\bar{q}_2q'_2$, ($l_i\neq e$).
The semi-leptonic {\tt CC10} process is of special interest for the
study of anomalous couplings since its final states offer the
most complete kinematical information for an experimental analysis of $W$ pair
production. 
 
Present experimental limits on anomalous couplings are not too stringent.
In the para\-meter space of $\alpha_{W\phi}$, $\alpha_W$, and $\alpha_{B\phi}$
the combined limits of LEP and D0 are \cite{Clare:1998aa} 
(see also ~\cite{Ellison:1998uy,Ellison:1998ub}) :
\begin{eqnarray}
\alpha_{W\phi}&=&
-0.03^{+0.06}_{-0.06},\nonumber\\
\alpha_W&=&
-0.03^{+0.08}_{-0.08},\vphantom{\frac{1^1}{1^1}}
\\
\alpha_{B\phi}&=&
-0.05^{+0.22}_{-0.20}.\nonumber
\end{eqnarray}
Here, the identities
\begin{eqnarray}
\alpha_{W\phi}&=&c_Ws_W\delta_Z,\\
\alpha_W&=&y_\gamma=\frac{s_W}{c_W}y_Z,\\
\alpha_{B\phi}&=&x_\gamma-c_Ws_W\delta_Z=-\frac{c_W}{s_W}
\left(x_Z+s_W^2\delta_Z\right)
\end{eqnarray}
are implied, and the anomalous couplings $x,y,\delta_Z$ are defined in
section~\ref{ano}.
These conventions are in accordance with
\cite{Gounaris:1996rz,Gounaris:1997}.

First signals of anomalous triple gauge boson couplings will be small if any.  
Since the total cross-section is not very sensitive to anomalous couplings, it 
is advantageous to study distributions.

The semi-analytical expressions of {\tt GENTLE} for 
total cross-sections with QED corrections in the standard model were derived
for the signal diagrams in \cite{Bardin:1993nb,Bardin:1996jw} and for the
background contributions in \cite{Bardin:1996uc,Bardin:1995vm}.
With the results presented in this article, {\tt GENTLE} 
may
be used also for predictions of $\mbox{d}\sigma/\mbox{d}\cos\theta$,
where $\theta$ is the production angle of one of the $W$ bosons.
 We present analytical expressions for the
{\it differential cross-section\/} for processes of the {\tt CC11} class in the
standard model in section \ref{SM} and 
the effects of {\it anomalous couplings\/} in section \ref{ano}.
In appendices we give some technical details of notations and the treatment of
QED corrections.
Numerical results are discussed in the corresponding sections.

The formulae of this article have been implemented in {\tt GENTLE}
version 2
\cite{Bardin:1996zz} which is currently  used for
experimental studies at LEP~2. 

\section{The Angular Distribution in the Standard Model}\label{SM}
\ezero

\subsection{The {\tt CC03} process}
\label{CC03}

The {\tt CC03} process is defined through reaction (\ref{cc03p}).
The fermion pairs $f_1\bar{f}'_1$ and $\bar{f}_2f'_2$  are the decay
products of $W^-$ and $W^+$:
\begin{equation}
	W^-\to d_1\bar{u'}_1,  \hspace{1cm} W^+\to \bar{d}_2u'_2,
\label{Wdecay}
\end{equation}
and have the invariant masses $s_1$ and $s_2$.
The scattering angle $\theta$ is defined as the angle between the
electron and the $W^-$ boson.

The differential cross-section may be written as a sum of $s$ and $t$-channel
contributions and of their interference 
 \cite{Muta:1986is}:
\begin{eqnarray}
  \frac{\mbox{d}\sigma_{\mbox{\scriptsize {\tt CC03}}}}{\mbox{d}\cos\theta}
  &=& \frac{\sqrt{\lambda}}{2\pi s^2}\int\mbox{d}s_1\mbox{d}s_2\,
  \left[{\cal C}^t
  {\cal G}^t(s;s_1,s_2,\cos\theta)+
  {\cal C}^s{\cal G}^s(s;s_1,s_2,\cos\theta)\right.
\nonumber\\&&\mbox{}
  \left.+{\cal C}^{st}{\cal G}^{st}(s;s_1,s_2,\cos\theta)
  \right] .
\label{cc03diff}
\end{eqnarray}
We give some notations, including the explicit expressions for the
{$\cal C$} and {$\cal G$} functions in appendix \ref{CC03results}.
\subsection{Contributions from background diagrams
\label{CC11}
}
We subdivide the background contributions into three parts:
\begin{equation}
  \frac{\mbox{d}\sigma_b}{\mbox{d}\cos\theta}=
      \frac{\mbox{d}\sigma_{sb}}{\mbox{d}\cos\theta}
      +\frac{\mbox{d}\sigma_{tb}}{\mbox{d}\cos\theta}
      +\frac{\mbox{d}\sigma_{bb}}{\mbox{d}\cos\theta}.
\label{diffback}
\end{equation}
The first term contains the interferences between the two $s$-channel
resonant diagrams and the eight background diagrams.
The second one describes the interferences between the $t$-channel exchange
diagram and background, and the third one the pure background
contributions.

We denote the various background diagrams by the type of
final state fermion coupling to the neutral gauge boson.
If e.g. an up-type anti-fermion couples to the photon or the $Z$
boson, we will call this a $u_1$-diagram.
In accordance with (\ref{Wdecay}), the subindex 1 (2) indicates by convention
that a 
fermion of the weak doublet with negative (positive) net charge is coupling to
the neutral boson.
We use the calculational method described in ~\cite{Bardin:1996uc} and
{\tt FORM}~\cite{Vermaseren:1991}. 
\subsubsection{Background--$s$-channel interference
\label{sback}}
There are~16~interferences between the two $s$-channel signal
diagrams and the eight background diagrams.
Each of these interferences is split up into two products
${\cal C}^{sa_i}_{+}{\cal G}^{sa_i}_{+}$ and
${\cal C}^{sa_i}_{-}{\cal G}^{sa_i}_{-}$:
\begin{equation}
  \frac{\mbox{d}\sigma_{sb}}
  {\mbox{d}\cos\theta}=\frac{\sqrt{\lambda}}{2\pi s^2}
  \int\mbox{d}s_1\mbox{d}s_2\,
  \sum\limits_{i=1,2}\sum\limits_{a=u,d}
  \left[{\cal C}^{sa_i}_{+}{\cal G}^{sa_i}_{+}+
{\cal C}^{sa_i}_{-}{\cal G}^{sa_i}_{-}\right].
\label{s-background}
\end{equation}
Summation index $a$ stands for up-type or down-type fermions of
doublet~$i$.
The coefficient functions are:
\begin{eqnarray}
  {\cal C}^{sa_i}_{\pm}&=&\sum\limits_{k,l=\gamma,Z}
  \frac{2}{(6\pi^2)^2}\mbox{Re}\frac{1}
  {D_{k}(s)D^*_{l}(s)D_W(s_1)D_W(s_2)D^*_W(s_{3-i})}
\nonumber\\&&\mbox{}
  \times g_{k}\left[L(e,k)L(e,l)\pm R(e,k)R(e,l)\right]
\nonumber\\&&\mbox{}
  \times L^2(F_1,W)L^2(F_2,W)L(f_a ^i,l)N_c(F_1)N_c(F_2).
\label{csfa}
\end{eqnarray}
The propagators are defined in (\ref{propag}) and the coupling constants
in (\ref{couplingconst}). 
The two independent kinematical functions for the 
$su_1$-interference are:
\begin{eqnarray}
{\cal G}_{-}^{su_1}(s,s_1,s_2)&=&
  \frac{3}{16}\frac{\cos\theta}{\sqrt{\lambda}}ss_2\left\{
  2s\left[s(s_1+s_2)-s_1^2-s_2^2\right]
  {\cal L}(s_1;s_2,s)+(s+s_1)^2-s_2^2\right\},
\nonumber\\
\\
{\cal G}_{+}^{su_1}(s,s_1,s_2)&=&
  \frac{3}{16}\frac{1-3\cos^2\theta}{\lambda}s^2s_1s_2
  \left[2ss_2{\cal L}(s_1;s_2,s)+s-s_1+s_2\right]
\nonumber\\&&\mbox{}
  -\frac{3ss_2}{16}\left[s(s_1+s_2)(1+\cos^2\theta)+2s_1s_2\sin^2\theta
  \right]{\cal L}(s_1;s_2,s)
\nonumber\\&&\mbox{}
  +\frac{ss_1}{8}(s_1-s-4s_2)
  +\frac{3s_2}{32}\left[s(3s_1-s_2-s)(1+\cos^2\theta)\right.
\nonumber\\ &&\mbox{}
  \left.\mbox{} + 2s_1(s_1-s_2)\sin^2\theta\right]
  +\frac{\lambda\sin^2\theta}{64}(s_1-s-s_2).
\label{gu1u1}
\end{eqnarray}

The logarithm
\begin{equation}
  {\cal L}(s;s_1,s_2)=\frac{1}{\sqrt{\lambda}}\ln
  \frac{s-s_1-s_2+\sqrt{\lambda}}{s-s_1-s_2-\sqrt{\lambda}}
\end{equation}
arises from integrating the fermion propagators in the background
diagrams.

The ${\cal G}^{sa_i}_-$-functions are proportional to  $\cos\theta$ and,
thus, they contribute only to the differential cross-section but do
not contribute to the total cross-section.
After integration over $\cos\theta$, (\ref{gu1u1}) yields 
eq.~(3.1) of \cite{Bardin:1996uc}.

One may obtain the $su_2$-interference by exchanging $s_1$ and $s_2$ in the
$su_1$-interference:
\begin{equation}
{\cal G}_{\pm}^{su_2}(s,s_1,s_2)=
{\cal G}_{\pm}^{su_1}(s,s_2,s_1).
\end{equation}

To construct the kinematical functions with the down-type fermion
coupling to the neutral vector boson, one may use the symmetry:
\begin{equation}
{\cal G}_{\pm}^{sd_1}(s,s_1,s_2)=
{\cal G}_{\pm}^{sd_2}(s,s_2,s_1)=
\mp{\cal G}_{\pm}^{su_1}(s,s_1,s_2).
\end{equation}

The coefficients of the ${\cal P}$ violating contributions in
(\ref{s-background}), ${\cal C}^{sa_i}_-$, vanish for pure photon exchange.


\subsubsection{Background--$t$-channel interference}
\label{tback}

The $t$-channel background interference is:
\begin{equation}
  \frac{\mbox{d}\sigma_{tb}}
  {\mbox{d}\cos\theta}=\frac{\sqrt{\lambda}}{2\pi s^2}
  \int\mbox{d}s_1\mbox{d}s_2\,
  \sum\limits_{i=1,2}\sum\limits_{a=u,d}
  {\cal C}^{ta_i}{\cal G}^{ta_i}.
\end{equation}
Due to the neutrino exchange in the $t$-channel,
only left-handed particles contribute and,
therefore, only one combination of couplings appears:
\begin{eqnarray}
  {\cal C}^{ta_i}&=&\sum\limits_{k=\gamma,Z}\frac{2}{(6\pi^2)^2}
  \mbox{Re}\frac{1}
  {D_W(s_1)D_W(s_2)D^*_{k}(s)D^*_W(s_{3-i})}
\nonumber\\&&\mbox{}
  \times L^2(E,W)L(e,k)L(f_i^a,k)L^2(F_1,W)L^2(F_2,W)N_c(F_1)N_c(F_2).
\end{eqnarray}

The kinematical functions are exceptionally asymmetric since
the integration over the fermion propagator in the background diagrams
is performed, while the neutrino propagator (\ref{nuprop}) in the
$t$-channel diagram is still present.
The kinematical function for the $tu_1$-interference is:
\begin{eqnarray}
\lefteqn{{\cal G}^{tu_1}(s,s_1,s_2) =}\nonumber\\
&&  \frac{-1}{\lambda}\left\{
  \frac{3}{4}\frac{\cos\theta}{\sqrt{\lambda}}\right.
  s^2s_1s_2^2(5\sin^2\theta-2)\left[\frac{1}{t_\nu}(s+s_1-s_2)
  +2s{\cal L}(s_1;s_2,s)\right]
\nonumber\\&&\mbox{}
  +\lambda\left[\frac{\sin^2\theta}{8t_\nu}[2s_1s_2(s_2-s_1)
  -6s^2s_2(s_1+s_2){\cal L}(s_1;s_2,s)-3ss_2(s+s_2)]\right.
\nonumber\\&&\mbox{}
  \left.+\frac{\sin^2\theta}{16}[(s-s_1)^2-s_2^2]+\frac{ss_1}{2}
  \vphantom{\frac{1}{t_\nu}}\right]
  +\frac{ss_1s_2}{t_\nu}\left[-\frac{3}{4}ss_2{\cal L}(s_1;s_2,s)
  (5s\sin^4\theta+4s_1\right.
\nonumber\\&&\mbox{}
  +4s_2)-\frac{1}{8}(3s_2^2-2ss_1+4s_1s_2-7s_1^2+30ss_2+9s^2)
  \sin^2\theta-\frac{1}{2}(3s_2^2-2s_1^2
\nonumber\\&&\mbox{}
  \left.-s_1s_2+2ss_1)\vphantom{\frac{3}{4}}\right]
  +\frac{3s^2s_2}{4}{\cal L}(s_1;s_2,s)([4s_1s_2+s_1^2+s_2^2-s(s_1+s_2)]
  \sin^2\theta
\nonumber\\&&\mbox{}
  -4[s_1s_2+s_1^2+s_2^2-s(s_1+s_2)])
  +\frac{ss_2\sin^2\theta}{8}(2s_1s_2-5s_1^2+3s_2^2
\nonumber\\&&\mbox{}
  -14ss_1 -3s^2)
  \left.+\frac{s}{2}(5s_1^2s_2-2s_1s_2^2-3s_2^3
  +5ss_1s_2+3s^2s_2)\right\}.
\label{tu1}
\end{eqnarray}

The expression for the $td_1$-interference becomes quite compact
using (\ref{tu1}):
\begin{eqnarray}
{\cal G}^{td_1}(s,s_1,s_2)&=&
  -{\cal G}^{tu_1}(s,s_1,s_2)
\nonumber\\&&\mbox{}
  -\frac{3ss_2}{\lambda}\left[\frac{\sin^2\theta}{4t_\nu}\left\{(s+s_1+s_2)
  [s_1(2s_1-s-s_2)-(s-s_2)^2]\right.\right.
\nonumber\\&&\mbox{}
  \left.-2\left[ss_1(s-s_1)^2+ss_2(s-s_2)^2
  +s_1s_2(s_1-s_2)^2\right]{\cal L}(s_1;s_2,s)\right\}
\nonumber\\&&\mbox{}
  \left.+s\left[s(s_1+s_2)-s_1^2-s_2^2\right]{\cal L}(s_1;s_2,s)
  +\frac{1}{2}\left[(s+s_1)^2-s_2^2\right]\right].
\label{td1}
\end{eqnarray}

The integral over $\cos\theta$ of (\ref{tu1}) yields
${\cal G}^{u,d}_{\tt CC11}$ and of (\ref{td1}) yields ${\cal
G}^{uu,dd}_{\tt CC11}$ (eq.~(3.12) in \cite{Bardin:1996uc}).

The remaining two ${\cal G}$ functions are easily constructed:
\begin{equation}
{\cal G}^{tu_2}(s,s_1,s_2)=
{\cal G}^{tu_1}(s,s_2,s_1)
\end{equation}
and
\begin{equation}
{\cal G}^{td_2}(s,s_1,s_2)=
{\cal G}^{td_1}(s,s_2,s_1).
\end{equation}


\subsubsection{Pure background}
\label{purebackground}

The pure background contribution is:
\begin{equation}
  \frac{\mbox{d}\sigma_{bb}}
  {\mbox{d}\cos\theta}=\frac{\sqrt{\lambda}}{2\pi s^2}
  \int\mbox{d}s_1\mbox{d}s_2\,
  \sum\limits_{a,b=u,d}\sum\limits_{i,j=1,2}
  \left[{\cal C}^{a_ib_j}_+{\cal G}^{a_ib_j}_++
{\cal C}^{a_ib_j}_-{\cal G}^{a_ib_j}_-\right].
\label{backcross}
\end{equation}

Again, we have to introduce additional coefficient functions ${\cal C}_-$
compared to the total cross-section, where only ${\cal C}_+$ functions
appear:
\begin{eqnarray}
  {\cal C}^{a_ib_j}_\pm&=&
  \sum\limits_{k,l=\gamma,Z}\frac{2}{(6\pi^2)^2}\mbox{Re}\frac{1}{D_{k}(s)
  D^*_{l}(s)D_W(s_{3-i})D^*_W(s_{3-j})}
\nonumber\\&&\mbox{}
  \times[L(e,k)L(e,l)\pm R(e,k)R(e,l)]
\nonumber\\&&\mbox{}
  \times L^2(F_1,W)L^2(F_2,W)N_c(F_1)N_c(F_2)
\nonumber\\&&\mbox{}
  \times L(f^a_i,k)L(f^b_j,l).
\label{cpb}
\end{eqnarray}

The potentially $2\times64$ kinematical functions in (\ref{backcross})
can be reduced to $2\times16$ functions in a first step since the
$\gamma$ and $Z$ exchange differ only in the coefficient functions~(\ref{cpb}).
With 
\begin{equation}
  {\cal G}^{a_ib_j}_\pm={\cal G}^{b_ja_i}_\pm  
\end{equation}
the number of independent ${\cal G}$-functions is further reduced
to $2\times10$.

Finally, we will need only five kinematical functions to
express them all.

The simplest cases are the squares of the various background diagrams
($a=b$ and $i=j$ in (\ref{backcross}));
they are given by:
\begin{eqnarray}
{\cal G}^{u_1u_1}_-(s,s_1,s_2)
&=&\frac{3}{4}\frac{\cos\theta}{\sqrt{\lambda}}ss_2\left\{
       \frac{1}{2}{\cal L}(s_1;s_2,s)\left[\left(s-s_1\right)^2
       -s_2^2\right]+s-s_1-s_2\right\}
\nonumber\\
\label{u1u1-}
\end{eqnarray}
and
\begin{eqnarray}
{\cal G}^{u_1u_1}_+(s,s_1,s_2)
&=&\frac{3}{8}\frac{1-3\cos^2\theta}{\lambda}
  ss_1s_2^2\left[
  {\cal L}(s_1;s_2,s)(s_2-s_1+s)+2\right]
  +\frac{1}{64}
  \lambda(1-\cos^2\theta)
\nonumber\\&&\mbox{}
  +\frac{3}{16}s_2(1+\cos^2\theta)\left[s{\cal L}(s_1;s_2,s)
  (s_1-s_2-s)-2s-s_1\right]
\nonumber\\&&\mbox{}
+\frac{1}{8}s_1(s+3s_2).
\label{u1u1+}
\end{eqnarray}
By integrating (\ref{u1u1+}) over $\cos\theta$ one gets 
${\cal G}^{ff}_{\tt CC11}$ in (3.3) of \cite{Bardin:1996uc} while
(\ref{u1u1-}) vanishes.
 
The other interferences between
background diagrams of the same doublet are:
\begin{eqnarray}
  {\cal G}^{u_2u_2}_\pm(s,s_1,s_2)&=&
  {\cal G}^{u_1u_1}_\pm(s,s_2,s_1),
\\
\nonumber\\
  {\cal G}^{d_1d_1}_\pm(s,s_1,s_2)&=&
  \pm{\cal G}^{u_1u_1}_\pm(s,s_1,s_2),\label{u1u1}
\\
\nonumber\\
  {\cal G}^{d_2d_2}_\pm(s,s_1,s_2)&=&
  \pm{\cal G}^{u_1u_1}_\pm(s,s_2,s_1).
\end{eqnarray}
With the aid of a neutral current function, one may prove the relation:
\begin{eqnarray}
{\cal G}^{u_1d_1}_+(s,s_1,s_2)
&=&
{\cal G}^{d_1d_1}_+(s,s_1,s_2)+
  {\cal G}^{u_1u_1}_+(s,s_1,s_2)-
  ss_2{\cal G}^{DD}_{422}(s,s_1,s_2) .
\label{arnd1}
\end{eqnarray}
The function ${\cal G}^{DD}_{422}(s,s_1,s_2)$ may be found in
appendix~\ref{NC20}.

The functions ${\cal G}^{a_ib_j}_-$ vanish in the neutral current
case and the analogue of (\ref{arnd1}) is: 
\begin{equation}
  {\cal G}^{u_1d_1}_-(s,s_1,s_2)=-\left[{\cal G}^{d_1d_1}_-(s,s_1,s_2)
  +{\cal G}^{u_1u_1}_-(s,s_1,s_2)\right] = 0.
  \label{analog1}
\end{equation}
The expressions for the other doublet are:
\begin{equation}
  {\cal G}^{u_2d_2}_\pm(s,s_1,s_2)=
  {\cal G}^{u_1d_1}_\pm(s,s_2,s_1).
\end{equation}

The interferences between diagrams from different doublets are more
complicated.
Here, we have
\begin{eqnarray}
{\cal G}^{u_1d_2}_-(s,s_1,s_2)&=&
  \frac{3}{8}\frac{\cos\theta}{\sqrt{\lambda}}s\left\{2s\left[s_2^2{\cal L}
  (s_1;s_2,s)-s_1^2{\cal L}(s_2;s,s_1)+\frac{s_2-s_1}{2}\right]
  -s_1^2+s_2^2\right\}\nonumber\\
\label{u1d2-}
\end{eqnarray}
%
and the lengthy expressions
\begin{eqnarray}
{\cal G}^{u_1d_2}_+(s,s_1,s_2)&=&
  -18\frac{s^2s_1^2s_2^2}{\lambda^3}(1+\sin^2\theta)
  s^2s_1s_2{\cal L}(s_1;s_2,s){\cal L}(s_2;s,s_1)
\nonumber\\&&\mbox{}
  -3s\left[s_1^2{\cal L}(s_2;s,s_1)+s_2^2{\cal L}(s_1;s_2,s)\right]
\nonumber\\&&\mbox{}
  \times\left[\frac{\sin^2\theta}{8}+\frac{s\cos^2\theta}{4\lambda}
  (s-\sigma)+\frac{s^2s_1s_2(1+\sin^2\theta)}{2\lambda^2}\left(
  2-3s\frac{s-3\sigma}{\lambda}\right)
  \right]
\nonumber\\&&\mbox{}
  -s(s_1-s_2)\left[s_1^2{\cal L}(s_2;s,s_1)
  -s_2^2{\cal L}(s_1;s_2,s)\right]
\nonumber\\&&\mbox{}
  \times\left[\frac{3\sin^2\theta}{8\lambda}(s-\sigma)+
  \frac{3ss_1s_2(1+\sin^2\theta)}{2\lambda^2}\left(
  1-3s\frac{s+\sigma}{\lambda}\right)\right]
\nonumber\\&&\mbox{}
  +\frac{3s^2s_1s_2(1+\sin^2\theta)}{4\lambda^2}\left[
  s^2-s_1^2-s_2^2-\frac{12ss_1s_2(s-\sigma)}{\lambda}\right]
\nonumber\\&&\mbox{}
  +\frac{s(1+\cos^2\theta)}{16\lambda}\left[4ss_1s_2+3(s_1^3+s_2^3)
  -(3s^2+7s_1s_2)\sigma\right]
\nonumber\\&&\mbox{}
  \vphantom{\frac{s^2_1}{\lambda^3}}
  -\frac{\sin^2\theta}{32}\left[\frac{24ss_1s_2(2s-\sigma)}{\lambda}
  +s^2-s_1^2-s_2^2-10s_1s_2\right]
\label{u1d2+}
\end{eqnarray}
and
\begin{eqnarray}
{\cal G}^{u_1u_2}_-(s,s_1,s_2)&\!=\!&
  -{\cal G}^{u_1d_2}_-(s,s_1,s_2)
  +\frac{s\cos\theta}{\sqrt{\lambda}}\left\{
  \frac{27s^2s_1^2s_2^2}{2\lambda^2}\biggl[s(\sigma-s)
  {\cal L}(s_1;s_2,s){\cal L}(s_2;s,s_1)\right.
\nonumber\\&&\mbox{}
  +(s_1-s-s_2){\cal L}(s_1;s_2,s)+(s_2-s-s_1){\cal L}(s_2;s,s_1)
  -2\biggr]
\nonumber\\&&\mbox{}
  +\frac{9ss_1s_2}{2\lambda}\left[\vphantom{\frac{4}{5}}s
  [3s_1s_2+s(\sigma-s)]{\cal L}(s_1;s_2,s){\cal L}(s_2;s,s_1)\right.
\nonumber\\&&\mbox{}
  +\left[s(s_1-s)+s_2\left(s_1-s_2-\frac{s}{2}\right)\right]
    {\cal L}(s_1;s_2,s)
\nonumber\\&&\mbox{}
  \left.+\left[s_2(s_2-s-s_1)-\frac{5}{2}ss_1\right]{\cal L}(s_2;s,s_1)
  -\frac{5}{4}(s+\sigma)\right]
\nonumber\\&&\mbox{}
  +\frac{3s_1}{4}
  \left[\vphantom{\frac{3}{2}}6s^2s_2{\cal L}(s_1;s_2,s)
  {\cal L}(s_2;s,s_1)+3ss_2{\cal L}(s_1;s_2,s)\right.
\nonumber\\&&\mbox{}
  \left.\left.-s(3s_2+2s_1){\cal L}(s_2;s,s_1)
  -\left( s+s_1 +\frac{3}{2}s_2 \right) \right] \right\}.
\label{u1u2-}
\end{eqnarray}
In (\ref{u1d2+}) and (\ref{u1u2-}) we use the abbreviation
\begin{equation}
\sigma=s_1+s_2.
\end{equation}
Further,
\begin{equation}
{\cal G}^{u_1u_2}_+(s,s_1,s_2)=
  \frac{1}{2}ss_1s_2{\cal G}^{DD}_{233}(\cos\theta,s,s_1,s_2)
  -{\cal G}^{u_1d_2}_+(s,s_1,s_2).
\label{arnd2}
\end{equation}
The neutral current function  ${\cal G}^{DD}_{233}(\cos\theta,s,s_1,s_2)$
can be found
in appendix~\ref{NC20}.
The integral of (\ref{u1d2+}) is
${\cal G}^{u,d}_{\tt CC11}$ defined in (3.10) of \cite{Bardin:1996uc}
and that of (\ref{u1d2-}) and (\ref{u1u2-}) vanish.

The remaining kinematical functions are
\begin{eqnarray}
{\cal G}^{d_1u_2}_\pm(s,s_1,s_2)&=&
{\cal G}^{u_1d_2}_\pm(s,s_2,s_1)
\end{eqnarray}
and
\begin{equation}
  {\cal G}^{d_1d_2}_\pm(s,s_1,s_2)=
  \pm{\cal G}^{u_1u_2}_\pm(s,s_1,s_2).
\end{equation}

\subsection{Numerical results
\label{numressm}}
Numerical results are obtained with the Fortran program
{\tt GENTLE} \cite{Bardin:1996zz}, version 2.02.
\\
QED initial state radiation (ISR) is treated as described in
appendix~\ref{qedcorrections} and in \cite{Bardin:1996zz}.

We use the numerical default input values, e.g.
$M_W = 80.230$ GeV, $\Gamma_W = 2.0855$ GeV, $M_Z = 91.1888$ GeV,
$\Gamma_Z = 2.4974$~GeV,
$\sin^2 \theta_{W} = 0.22591$, $\alpha_{em} = 1/137.0359895$,
$\alpha_s = 0.12 $, no Cabibbo mixing, and the {\tt GENTLE} flag settings
\[
\begin{array}{c@{\mbox{}=\mbox{}}c@{\mbox{}=\mbox{}}c
@{\mbox{}=\mbox{}}c@{\mbox{}=\mbox{}}c@{\mbox{}=\mbox{}}c
@{\mbox{}=\mbox{}}c@{\mbox{}=\mbox{}}c}
 {\tt IPROC} &{\tt IINPT} &{\tt IONSHL} &{\tt IZETTA}
  &\multicolumn{4}{l}{ 1}\\
{\tt ICONVL} &{\tt IIQCD} &{\tt IDCS} &{\tt IMAP} &{\tt IRSTP}
&{\tt IMMIN} &{\tt IMMAX} & 1\\
{\tt IGAMWS} &{\tt IGAMZS}
&\multicolumn{6}{l}{ 1}\\
{\tt IGAMW} &{\tt ITNONU} &{\tt IQEDHS} &{\tt ICOLMB} &{\tt IZERO}
&{\tt IBIN} &{\tt IRMAX }& 0.
\end{array}
\]
The flags {\tt IBORNF}, {\tt IBCKGR}, {\tt ICHNNL} are varied in
an obvious
way. 
For calculations within the standard model {\tt IANO} is set equal to 0. 
The flags {\tt IGAMWS} and {\tt IGAMZS} are chosen such that the boson 
widths are taken to be constant.
For the related problems with gauge invariance see
\cite{Argyres:1995ym,Beenakker:1996kn,Beenakker:1996kt}.

\begin{figure}
  \begin{center}
    \epsfxsize=15.5cm
    \leavevmode
    \epsffile{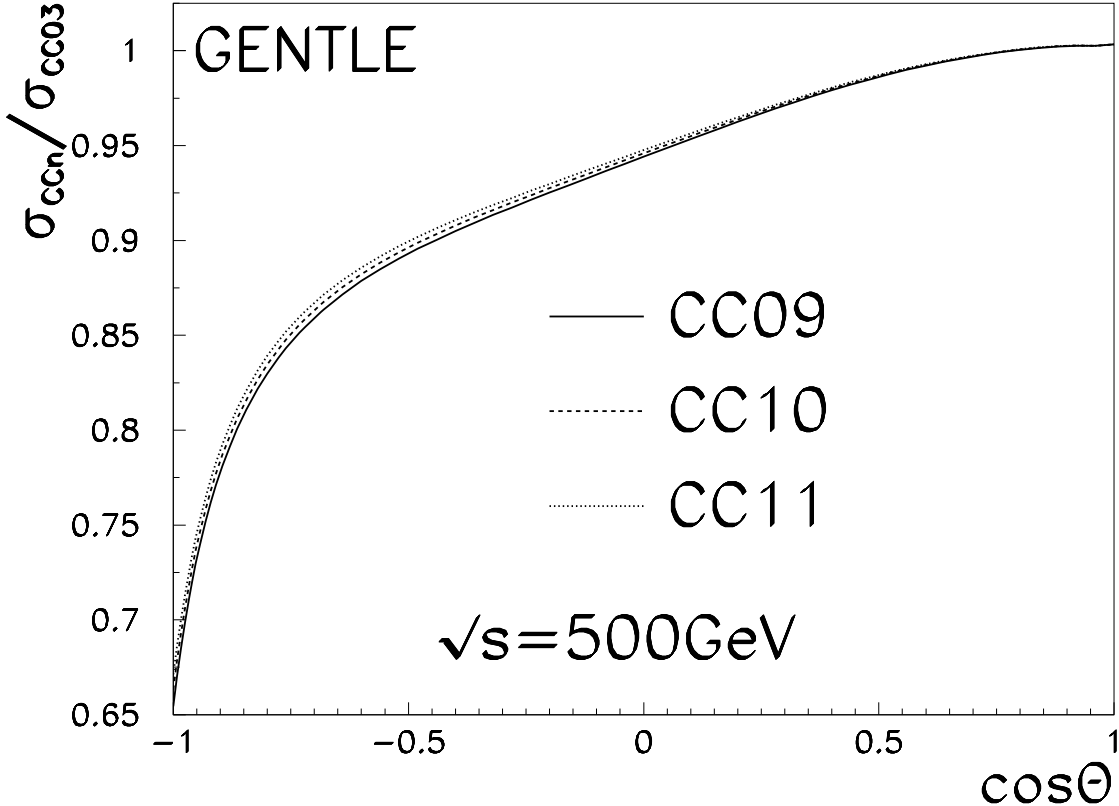}
  \end{center}
\vspace{-1.0cm}
\caption{\label{background}\it The ratios {\tt CC09/CC03},
{\tt CC10/CC03}, and {\tt CC11/CC03} without QED corrections.}
\end{figure}

In figure~\ref{background} the net size of the background effects 
at a center-of-mass energy of 500~GeV is shown
as the ratio of the signal plus background cross-section to the signal cross-section 
\begin{equation}
R=\frac{\mbox{d}\sigma_{\mbox{\scriptsize{\tt CC11}}}/\mbox{d}\cos\theta}
{\mbox{d}\sigma_{\mbox{\scriptsize \tt CC03}}/\mbox{d}\cos\theta}.
\end{equation}
The background contributions may become sizeable for large scattering angles.
For extreme backward production, $\cos\theta=-1$, the effect is
larger than 30\%.
At $\sqrt{s} = 190$ GeV, $\mbox{d}\sigma_b/\mbox{d}\cos\theta$
(see (\ref{diffback}) is less than
0.3\% of $\sigma_{\tt CC03}$ in the whole range of the scattering angle;
for more details see~\cite{Biebel:1997id}.

In tables~\ref{tabvalborn} and \ref{tabvalqed} we present numerical
data which may be of some use for precision comparisons of different
numerical programs. 
For this purpose, we ran {\tt GENTLE} at high numerical precision but
still a reasonable computing time at a PC with a Pentium~133~MHz
processor.
The numerical reliability was controlled by varying the parameters
$\epsilon$ (for the relative error of Simpson integration) and
$\delta$ (a technical cut parameter improving numerical stability in
some edges of the phase space)  in {\tt GENTLE}.  
The {\em technical uncertainties} in the
last digits shown in the tables 
are of the order one or smaller.  
Without ISR the cross-sections are obtained with 
$\epsilon=10^{-8}$ and $\delta=10^{-5}$.
With ISR, the corresponding values are 
$\epsilon=3 \times 10^{-5}$ and $\delta=10^{-4}$ (if {\tt IBCKGR}=1
and {\tt IBORNF}=1, then $\delta$ = $10^{-3}$).

\begin{figure}[ht]
  \begin{center}
    \epsfxsize=15.5cm
    \leavevmode
    \epsffile{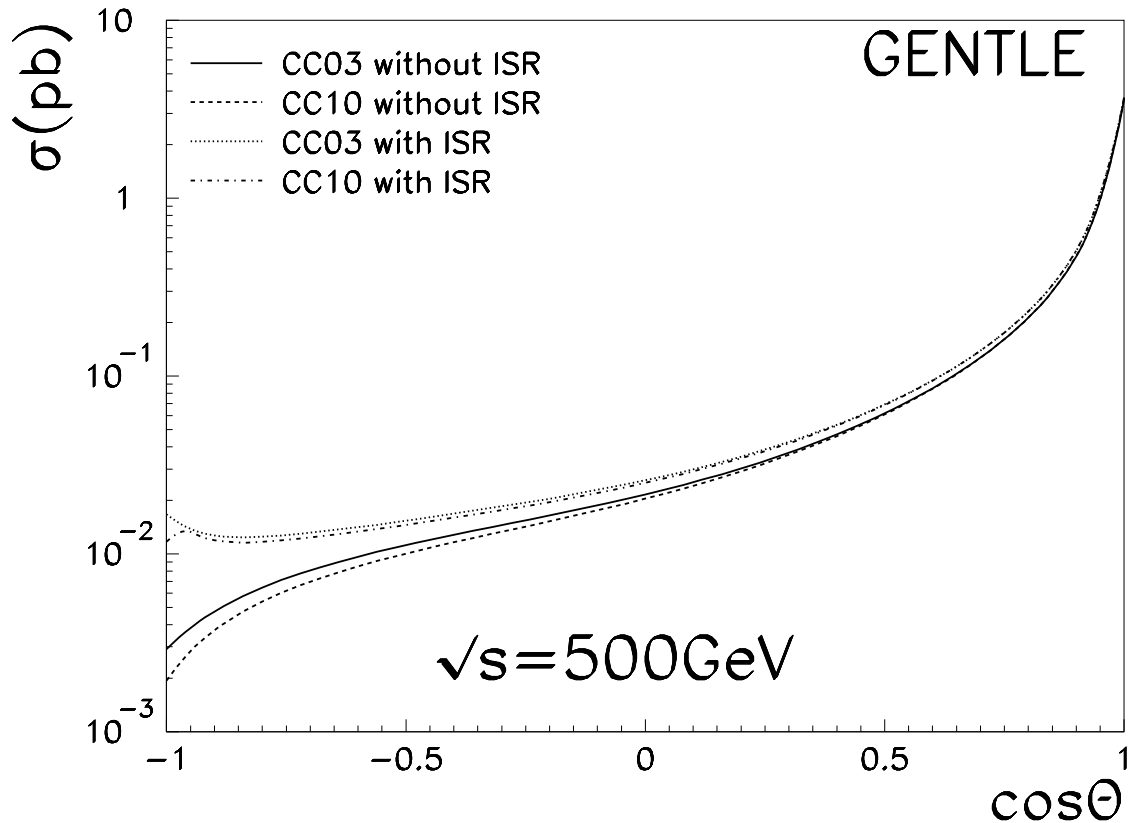}
  \end{center}
  \caption{\label{qed.ps}\it Differential cross-section for $e^+e^-\to
  \mu^-\bar{\nu}_\mu u\bar{d}$ with various corrections.}
\end{figure}

Although there are different coupling constants and even a different
number of contributing diagrams for the various {\tt CC09, CC10, CC11}
processes, one observes only small differences between the various
cross-sections, especially at LEP~2 energies.
One may suspect that this is due to certain relations between the 
relevant coupling constant combinations in the cross-sections which
make the latter being dependent on only weak iso-spins.
For example, the $s$-channel background interference
contributions
$
{\cal C}_+^{su_i}{\cal G}_+^{su_i}+
{\cal C}_+^{sd_i}{\cal G}_+^{sd_i}
$
are identical for all processes of the {\tt CC11} class.
Here, only the parity violating terms lead to different background effects
in this interference.
Similarly, for the $t$-channel background interference the flavour
depending combination ${\cal G}^{tu_i}+{\cal G}^{td_i}$ is suppressed
by cancellations as it can be seen in (\ref{td1}).
Obviously, the effects which are similar for all final states of the {\tt CC11}
class are the numerically dominating background corrections.  

QED corrections are shown in figure~\ref{qed.ps}
at $\sqrt{s}=500$ GeV.
There is a considerable cross-section enhancement for $\cos\theta<-0.5$. 
The background effect in this region of $\cos\theta$ is reduced, but still
sizeable.

\begin{table}[ht]
\begin{center}
\begin{tabular}{|c|r@{.}l|r@{.}l|r@{.}l|r@{.}l|r@{.}l|}
\hline
 $\sqrt{s}$ (GeV)&\multicolumn{2}{|c|}{$\cos\theta$}&
 \multicolumn{2}{|c|}{$\sigma_{\tt CC03}$ (pb)}&
 \multicolumn{2}{|c|}{$\sigma_{\tt CC09}$ (pb)} &
  \multicolumn{2}{|c|}{$\sigma_{\tt CC10}$ (pb)}&
  \multicolumn{2}{|c|}{$\sigma_{\tt CC11}$ (pb)}\\\hline
       &--0&8&0&0943803&0&0303466&0&0945159&0&294360\\
\cline{2-11}
190 &0&0&0&216791&0&0697497&0&217241&0&676600\\
\cline{2-11}
       &0&8&0&790432&0&253951&0&791012&2&46385\\\hline
       &--0&8&0&00646155&0&00172150&0&00539258&0&0168880\\
\cline{2-11}
500 &0&0&0&0215908&0&00654501&0&0204228&0&0637232\\
\cline{2-11}
       &0&8&0&212368&0&0682179&0&212503&0&661957\\\hline
       &--0&8&0&00244188&0&000403725&0&00128268&0&00407275\\
\cline{2-11}
1000 &0&0&0&00614750&0&00155423&0&00484706&0&0151154\\
\cline{2-11}
       &0&8&0&0535172&0&0170530&0&0530947&0&165310\\\hline
\end{tabular}
\end{center}
\caption{\it Differential cross-sections
  without ISR. The {\tt CC03} cross-section is calculated with the
  branching ratios for the {\tt CC10} process.
\label{tabvalborn}}
\end{table}

\begin{table}[ht]
\begin{center}
\begin{tabular}{|c|r@{.}l|r@{.}l|r@{.}l|r@{.}l|r@{.}l|}
\hline
 $\sqrt{s}$ (GeV)&\multicolumn{2}{|c|}{$\cos\theta$}&
 \multicolumn{2}{|c|}{$\sigma_{\tt CC03}$ (pb)}&
 \multicolumn{2}{|c|}{$\sigma_{\tt CC09}$ (pb)} &
  \multicolumn{2}{|c|}{$\sigma_{\tt CC10}$ (pb)}&
  \multicolumn{2}{|c|}{$\sigma_{\tt CC11}$ (pb)}\\\hline
       &--0&8&0&09064&0&02915&0&09077&0&2827\\
\cline{2-11}
190 &0&0&0&1971&0&06340&0&1975&0&6150\\
\cline{2-11}
       &0&8&0&6868&0&2206&0&6871&2&140\\\hline
       &--0&8&0&01248&0&003749&0&01170&0&0365\\
\cline{2-11}
500 &0&0&0&02590&0&008036&0&02506&0&0782\\
\cline{2-11}
       &0&8&0&2308&0&07418&0&2311&0&7198\\\hline
       &--0&8&0&005282&0&001441&0&004510&0&01412\\
\cline{2-11}
1000 &0&0&0&007757&0&00223&0&006939&0&02163\\
\cline{2-11}
       &0&8&0&0613&0&01963&0&06114&0&1904\\\hline
\end{tabular}
\end{center}
\caption{\it Differential cross-sections
  with ISR. The {\tt CC03} cross-section is calculated with the
  branching ratios for the {\tt CC10} process.
\label{tabvalqed}}
\end{table}


\section{Anomalous Couplings}
\label{ano}
\ezero

We now extend the Lagrangian of the standard model by anomalous
triple gauge boson couplings.
We allow terms that obey Lorentz invariance and ${\cal CP}$ invariance.
In addition, for the electromagnetic interaction we forbid ${\cal C}$
or ${\cal P}$
violation and will not modify its strength.

%
%
These conditions are fulfilled by the Lagrangian proposed in
\cite{Berends:1995dn}:
\begin{eqnarray}
  {\cal  L}&=&
  -ie\left[A_\mu\left(W^{-\mu\nu}W^+_\nu-W^{+\mu\nu}W^-_\nu\right) 
  +F_{\mu\nu}W^{+\mu}W^{-\nu}\right]-iex_\gamma
  F_{\mu\nu}W^{+\mu}W^{-\nu}
\nonumber\\&&\mbox{}
  -ie\cot\Theta_w\left[Z_\mu\left(W^{-\mu\nu}W^+_\nu-W^{+\mu\nu}W^-_\nu\right) 
  +Z_{\mu\nu}W^{+\mu}W^{-\nu}\right]-iex_ZZ_{\mu\nu}W^{+\mu}W^{-\nu}
\nonumber\\&&\mbox{}
  -ie\delta_Z\left[Z_\mu\left(W^{-\mu\nu}W^+_\nu
  -W^{+\mu\nu}W^-_\nu\right)+Z_{\mu\nu}W^{+\mu}W^{-\nu}\right]
\nonumber\\&&\mbox{}
  -ie\frac{y_\gamma}{M_W^2}F^{\nu\lambda}W^-_{\lambda\mu}W^{+\mu}_\nu
  -ie\frac{y_Z}{M_W^2}Z^{\nu\lambda}W^-_{\lambda\mu}W^{+\mu}_\nu
\nonumber\\&&\mbox{}
  +\frac{ez_Z}{M_W^2}\partial_\alpha\tilde{Z}_{\rho\sigma}\left(
  \partial^\rho W^{-\sigma}W^{+\alpha}-\partial^\rho
  W^{-\alpha}W^{+\sigma}+\partial^\rho
  W^{+\sigma}W^{-\alpha}-\partial^\rho W^{+\alpha}W^{-\sigma}\right).
\end{eqnarray}

The term with the dual field tensor $\tilde{Z}$ violates both
${\cal C}$ and ${\cal P}$.
With a multipole expansion, one gets the electromagnetic charge $Q_W$, the
magnetic dipole moment $\mu_W$, and the electric quadrupole moment
$q_W$~\cite{Aronson:1969aa,Gounaris:1996rz}:
\begin{eqnarray}
Q_W&=&e,\\
\mu_W&=&\frac{e}{2M_W}(2+x_\gamma+y_\gamma),\\
q_W&=&-\frac{e}{M^2_W}(1+x_\gamma-y_\gamma).
\end{eqnarray}

The anomalous couplings $x_\gamma$, $x_Z$, $y_\gamma$, $y_Z$, $z_Z$,
and $\delta_Z$ produce additional contributions to the
cross-section of $W$ pair production.
The largest contributions will come from resonant diagrams
(section \ref{anoCC03}), but others
are also coming from the interference between anomalous $s$-channel
signal diagrams and background (section \ref{anoCC11}).


\subsection{Anomalous contributions to the {\tt CC03} process}
\label{anoCC03}

We write the cross-section for doubly resonant scattering
with anomalous couplings in the following form:
\begin{eqnarray}
  \frac{\mbox{d}\sigma^{\mbox{\scriptsize ano}}_{\mbox{\scriptsize CC03}}}
  {\mbox{d}\cos\theta}&=&
  \frac{\sqrt{\lambda}}{2\pi s^2}
  \int\mbox{d}s_1\mbox{d}s_2\,\left[\sum\limits_{nm}
  {\cal C}^s_{nm}{\cal G}^s_{nm}(s;s_1,s_2,\cos\theta)
  +\sum\limits_n{\cal C}^{st}_n{\cal G}^{st}_n(s;s_1,s_2,\cos\theta)
  \right].\nonumber\\
\label{anores}
\end{eqnarray}
The sums over $n$,$m$ run over $x$, $y$, $\delta$, and $z$ and the
standard model couplings.
The first sum in (\ref{anores}) describes the $s$-channel interferences 
and the second sum the anomalous $st$-interferences.
The coefficient functions are:
\begin{eqnarray}
  {\cal C}^{s}_{nm}&=&\sum\limits_{k,l=\gamma,Z}\frac{2}
  {(6\pi^2)^2}\mbox{Re}
  \frac{1}{|D_W(s_1)|^2|D_W(s_2)|^2D_{k}(s)D^*_{l}(s)}
\nonumber\\&&\mbox{}
  \times g^n_{k}g_{l}^m[1+(1-\delta^n_m)\delta^k_l]
  A^{nm}_{kl}
\nonumber\\&&\mbox{}
  \times L^2(F_1,W)L^2(F_2,W)N_c(F_1)N_c(F_2),
\label{Csnm}
\\
  {\cal C}^{st}_{n}&=&\sum\limits_{k=\gamma,Z}\frac{2}
  {(6\pi^2)^2}\mbox{Re}
  \frac{1}{|D_W(s_1)|^2|D_W(s_2)|^2D_{k}(s)}
\nonumber\\&&\mbox{}
  \times g_k^n L(e,l)L^2(F_1,W)L^2(F_2,W)L^2(e,W)
  N_c(F_1)N_c(F_2),
\end{eqnarray}
with 
\begin{equation}
  \begin{array}{rclrcl}
    g_\gamma^x&=&gs_Wx_\gamma,\mbox{\hspace{1cm}}&
    g_Z^x&=&gs_Wx_Z,
\\
\\
    g_\gamma^y&=&\frac{\displaystyle gs_Wy_\gamma}{\displaystyle
    M_W^2},&
    g_Z^y&=&\frac{\displaystyle gs_Wy_Z}{\displaystyle M_W^2},
\\
\\
    g_Z^\delta&=&gs_W\delta_Z,&g_Z^z&=&\frac{\displaystyle gs_Wz_Z}
    {\displaystyle M_W^2}.
  \end{array}
\label{couplings}
\end{equation}
and the standard model couplings
\begin{equation}
  \begin{array}{rclrcl}
    g_\gamma^{\rm SM}&=&gs_W,\mbox{\hspace{1cm}}&
    g_Z^{\rm SM}&=&gc_W.
  \end{array}
\label{smcouplings}
\end{equation}
The constant $A^{nm}_{kl}$ is defined as follows:
\begin{eqnarray}
A^{zm}_{kl}=A^{mz}_{kl}&=&L(e,k)L(e,l)-R(e,k)R(e,l)\mbox{\hspace{.5cm}
for } m\neq z\\
A^{nm}_{kl}&=&L(e,k)L(e,l)+R(e,k)R(e,l)\mbox{\hspace{.5cm} otherwise}
\end{eqnarray}

The $\delta^l_k$ in (\ref{Csnm}) is the Kronecker symbol.
Note that the pure standard model contributions are already treated
in section \ref{CC03} and should not be counted twice.

The anomalous kinematic functions in eq.~(\ref{anores}) are for the
$st$-interference:
\begin{eqnarray}
  {\cal G}^{st}_x&=&\frac{1}{8}s\left[(s_1+s_2)\left(s-s_1-s_2-\frac
  {2s_1s_2}{t_\nu}\right)+\frac{\lambda}{4}\sin^2\theta\right],
\\
\nonumber\\
  {\cal G}^{st}_y&=&\frac{1}{4}ss_1s_2
  \left[s-s_1-s_2-\frac{2s_1s_2}{t_\nu}\right],
\\
\nonumber\\
  {\cal G}^{st}_z&=&\frac{1}{16}\lambda s\left[2(s_1+s_2)
  -\frac{\sin^2\theta}{t_\nu}(s_1(s-s_1)+s_2(s-s_2))\right],
\end{eqnarray}
while for the $s$-channel contributions:
\begin{eqnarray}
  {\cal G}^s_{xx}& =&\frac{1}{128}\lambda s\left[(s_1+s_2)
  (1+\cos^2\theta)+s\sin^2\theta\right],
\\
\nonumber\\
  {\cal G}_{xy}^s& =&\frac{1}{64}\lambda ss_1s_2(1+\cos^2\theta),
\\
\nonumber\\
  {\cal G}^s_{sx}={\cal G}^s_{x\delta}&=&\frac{1}{128}\lambda
  s\left[4(s_1+s_2)+(s-s_1-s_2)\sin^2\theta\right],
\\
\nonumber\\
  {\cal  G}^s_{yy}&=&\frac{1}{128}\lambda ss_1s_2
  \left[2s\sin^2\theta+(s_1+s_2)(1+\cos^2\theta)\right],
\\
\nonumber\\
  {\cal G}^s_{sy}={\cal G}^s_{y\delta}&=&\frac{1}{16}\lambda ss_1s_2,
\\
\nonumber\\
  {\cal G}^s_{zz}&=&\frac{1}{128}\lambda^2s(s_1+s_2)
  (1+\cos^2\theta),
 \\
\nonumber\\
  {\cal G}^s_{s\delta}={\cal G}^s_{\delta\delta}&=&\frac{1}{32}\lambda
  \left[2s(s_1+s_2)+\left(3s_1s_2+\frac{\lambda}{4}\right)\sin^2
  \theta\right],
\\
\nonumber\\
  {\cal G}^s_{xz}&=& \frac{1}{64}\lambda^{\frac{3}{2}}
  s(s_1+s_2)\cos\theta,
\\
\nonumber\\
  {\cal G}^s_{yz}&=&\frac{1}{32}\lambda^{\frac{3}{2}}
  ss_1s_2\cos\theta,
\\
\nonumber\\
  {\cal G}^s_{sz}={\cal G}^s_{z\delta}&=&-\frac{1}{32}\lambda^{\frac{3}{2}}
  s(s_1+s_2)\cos\theta.
\end{eqnarray}


\subsection{Interferences of anomalous contributions with background}
\label{anoCC11}

Finally, we treat interferences of the anomalous $s$-channel diagrams
with background:
\begin{equation}
  \frac{\mbox{d}\sigma_{sb}^{\mbox{\scriptsize ano}}}
  {\mbox{d}\cos\theta}=\frac{\sqrt{\lambda}}{2\pi s^2}
  \int\mbox{d}s_1\mbox{d}s_2\,
  \sum\limits_{a=u,d}\sum\limits_{i=1,2}\sum\limits_n
  \left[{\cal C}^{sa_i}_{+,n}{\cal G}^{sa_i}_{+,n}+
{\cal C}^{sa_i}_{-,n}{\cal G}^{sa_i}_{-,n}\right].
\label{s-back-ano}
\end{equation}
The  coefficient functions are
\begin{eqnarray}
  {\cal C}^{sa_i}_{\pm,n}&=&\sum\limits_{k,l=\gamma,Z}
  \frac{2}{(6\pi^2)^2}\mbox{Re}\frac{1}
  {D_{k}(s)D^*_{l}(s)D_W(s_1)D_W(s_2)D^*_W(s_{3-i})}
\nonumber\\&&\mbox{}
  \times g^n_{k}\left[L(e,k)L(e,l)\pm R(e,k)R(e,l)\right]
\nonumber\\&&\mbox{}
  \times L^2(F_1,W)L^2(F_2,W)L(f_a ^i,l)N_c(F_1)N_c(F_2),
\end{eqnarray}
where $g^n_{k}$ stands for all couplings given in
eq.~(\ref{couplings}).

The anomalous kinematical functions are:
\begin{eqnarray}
{\cal G}_{-,x}^{su_1}(s,s_1,s_2)&=&
  \frac{3}{32}\frac{\cos\theta}{\sqrt{\lambda}}ss_2
  \left\{2s[s(s_1+s_2)-s_1^2-s_2^2]{\cal L}(s_1;s_2,s)
  +(s+s_1)^2-s_2^2\right\},
\nonumber\\
\\
{\cal G}_{+,x}^{su_1}(s,s_1,s_2)&=&
  \frac{3}{32}\frac{1-3\cos^2\theta}{\lambda}s^2s_1s_2
  \left[2ss_2{\cal L}(s_1;s_2,s)+s-s_1+s_2\right]
\nonumber\\&&\mbox{}
  -\frac{3s}{32}(1+\cos^2\theta)ss_2(s_1+s_2){\cal L}(s_1;s_2,s)
\nonumber\\&&\mbox{}
  +\frac{ss_2}{64}(1-3\cos^2\theta)(s+s_2-s_1)
\nonumber\\&&\mbox{}
  +\frac{s}{16}\left[s_1^2-s_2^2-s(s_1+s_2)
  -\frac{\lambda\sin^2\theta}{4}\right],
\\
\nonumber\\
{\cal G}_{-,y}^{su_1}(s,s_1,s_2)&=&
  \frac{3}{32}\frac{\cos\theta}{\sqrt{\lambda}}ss_1s_2
  \left\{2ss_2[2s-(s_1+s_2)]{\cal L}(s_1;s_2,s)
  -2s_2^2+2s_1s_2\right.
\nonumber\\&&\mbox{}
  \left.\vphantom{ss_2(2s-(s_1+s_2)){\cal L}}+3ss_2-ss_1+s^2\right\},
\\
{\cal G}_{+,y}^{su_1}(s,s_1,s_2)&=&
  \frac{ss_1s_2}{64}\left\{6\frac{1-3\cos^2\theta}{\lambda}ss_2
  \left\{s[s-(s_1+s_2)]{\cal L}(s_1;s_2,s)+s+s_1-s_2\right\}\right.
\nonumber\\&&\mbox{}
  +(1-3\cos^2\theta)[s-2ss_2{\cal L}(s_1;s_2,s)]
  -16ss_2{\cal L}(s_1;s_2,s)
\nonumber\\&&\mbox{}
  \left.\mbox{} -8(s-s_1+s_2)
  \vphantom{\frac{3s^2s_1s_2(1-3\cos^2\theta)}{32\lambda}}\right\},
\\
{\cal G}_{-,z}^{su_1}(s,s_1,s_2)&=&
  \frac{1}{32}\frac{\cos\theta}{\sqrt{\lambda}} 
  s\left\{6ss_1s_2[2ss_2{\cal L}(s_1;s_2,s)+s-s_1+s_2]
  \vphantom{s_1^2}\right.
\nonumber\\&&\mbox{}
  \left.+\lambda\left[6ss_2(s_1+s_2){\cal L}(s_1;s_2,s)
  +s(2s_1+3s_2)-s_1s_2-2s_1^2
  +3s_2^2\right]\right\},
\nonumber\\
\\
{\cal G}_{+,z}^{su_1}(s,s_1,s_2)&=&
  \frac{3}{64}(1+\cos^2\theta)ss_2\left\{2s[s_1^2+s_2^2-s(s_1+s_2)]
  {\cal L}(s_1;s_2,s)\right.
\nonumber\\&&\mbox{}
  \left.+s_2^2-(s+s_1)^2\right\}
  \vphantom{\frac{\cos\theta}{\sqrt{\lambda}}},
\\
{\cal G}_{-,\delta}^{su_1}(s,s_1,s_2)&=&{\cal G}_{-}^{su_1}
(s,s_1,s_2),
\\
{\cal G}_{+,\delta}^{su_1}(s,s_1,s_2)&=&{\cal G}_{+}^{su_1}
(s,s_1,s_2).
\end{eqnarray}

The remaining kinematical functions can be calculated with the equations
\begin{equation}
{\cal G}_{\pm,a}^{sd_1}(s,s_1,s_2)=
{\cal G}_{\pm,a}^{sd_2}(s,s_2,s_1)=
\pm{\cal G}_{\pm,a}^{su_2}(s,s_2,s_1)=
\pm{\cal G}_{\pm,a}^{su_1}(s,s_1,s_2),
\end{equation}
\begin{equation}
{\cal G}_{\pm,z}^{sd_1}(s,s_1,s_2)=
{\cal G}_{\pm,z}^{sd_2}(s,s_2,s_1)=
\mp{\cal G}_{\pm,z}^{su_2}(s,s_2,s_1)=
\mp{\cal G}_{\pm,z}^{su_1}(s,s_1,s_2),
\end{equation}
where $a$ stands for $x_\gamma$, $x_Z$, $y_\gamma$, $y_Z$, and
$\delta_Z$.



\subsection{Numerical results}

\begin{figure}
  \begin{center}
    \epsfxsize=15.5cm
    \leavevmode
    \epsffile{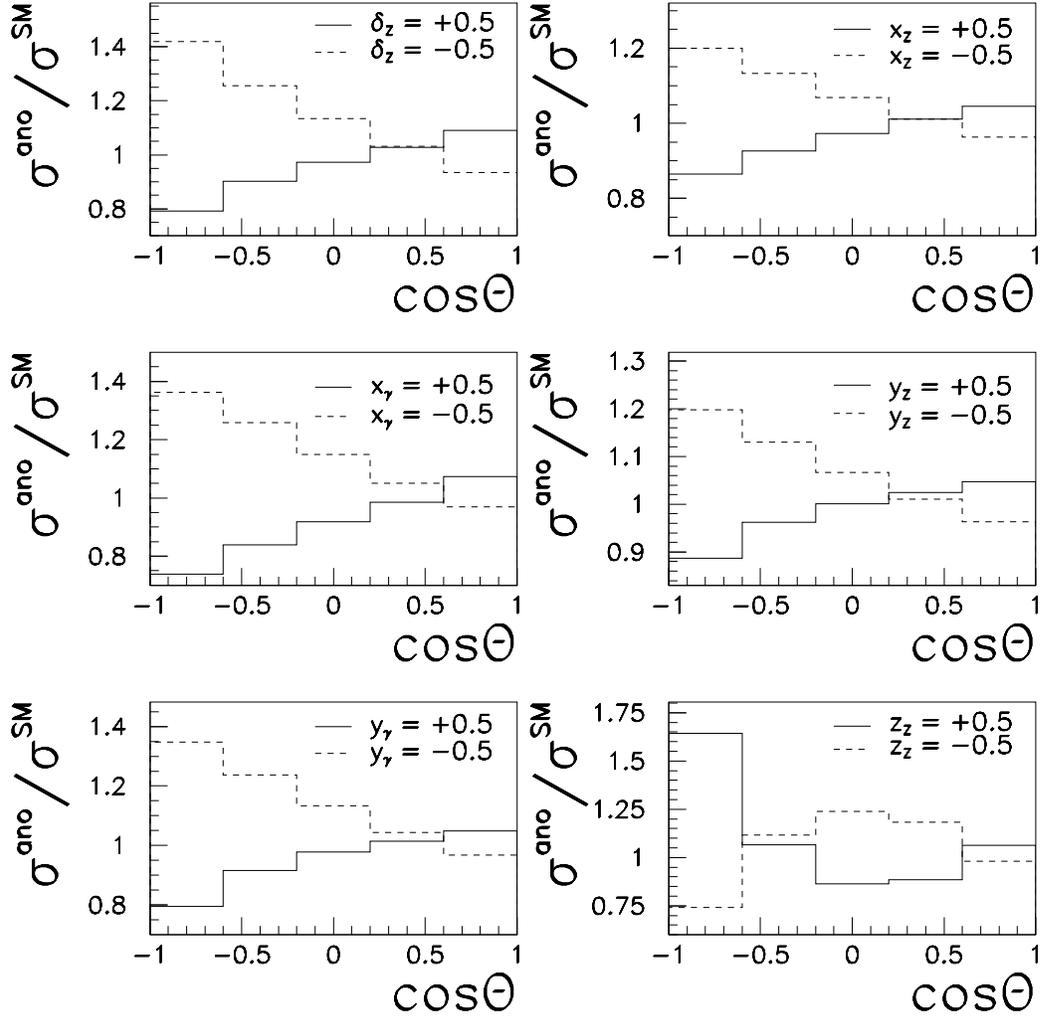}
  \end{center}
\caption{\label{anoall}\it The ratio of cross-sections with anomalous
  couplings to standard model cross-sections without background and
  without ISR corrections at $\sqrt{s}=190$ {\rm GeV}.
  In each figure only one anomalous coupling differs from~zero.}
\end{figure}

In figure \ref{anoall} we show the bin-integrated differential
cross-section for all the six anomalous couplings at
190~GeV.
In each case only one anomalous coupling is allowed to differ from
zero.
The figure is in excellent agreement
with an analogous figure in \cite{Berends:1995dn}.
Comparisons with the Monte Carlo event generator
{\tt WOPPER} \cite{Anlauf:1994sc,Anlauf:1996wq} show also 
agreement within the statistical accuracy of the MC program.

As an application, we shortly describe a study on the discriminative
power of $W$ pair production 
with respect to parity conserving and violating 
anomalous triple boson couplings.
At $\sqrt{s} = 500$ GeV with an integrated luminosity ${\cal L} =
50$ fb$^{-1}$, about 80~000 semi-leptonic $W$ pair decays are
produced.  
The anomalous couplings appear in the cross-section at most
bilinearly.
Allowing e.g. for two anomalous couplings $A$ and $B$ simultaneously,
one may use the ansatz:    
\begin{eqnarray}
\sigma_{theor} &=& \sigma^{\mbox{\scriptsize SM}} + A  \sigma_1 +
      A^2 \sigma_{11}
      +B \sigma_{2} + B^2 \sigma_{22} +AB \sigma_{12}.
\label{2fl}
\end{eqnarray}
After having calculated $\sigma^{\mbox{\scriptsize SM}},
\sigma_1, \sigma_{11}$, \dots
with {\tt
  GENTLE} (or another program) within a given model and for definite
experimental conditions, one may confront experimental data 
with the predictions.
For a study of sensitivities, we use $\sigma_{theor}$ for the
simulation of 
$\sigma_{meas} \pm \sqrt{\sigma_{meas}/(6{\cal L})}$, the assumed
measured  
cross-section with $1\sigma$ deviations of the counting rates for the
sum of all six semi-leptonic production channels.
For definiteness we use for $\sigma_{meas}$ the standard model
prediction $\sigma^{\mbox{\scriptsize SM}}$ and apply no experimental
cuts.
The solutions of eq.~(\ref{2fl}) for $A$ and $B$ are ellipses in the
plane.
Allowed pairs of coupling values are located in the area between
the two limiting ellipses.
\begin{figure}
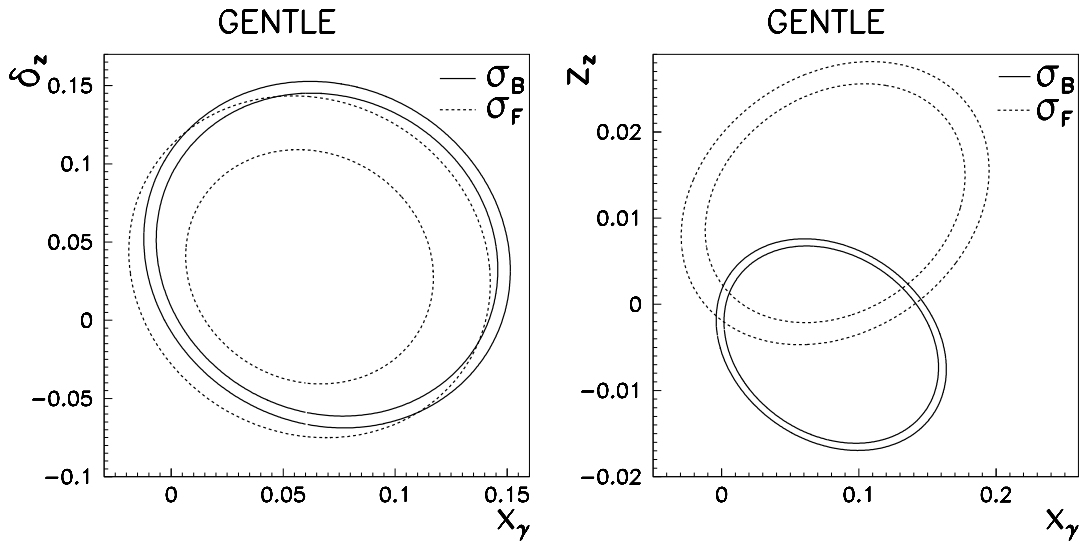

  \begin{center}
    \epsfxsize=7.3cm
    \leavevmode
    \epsffile{fig.4a}
    \epsfxsize=7.3cm
    \leavevmode
    \epsffile{fig.4b}
  \end{center}
\begin{center}
\caption{\label{rings}\it $1\sigma$-bounds at 500~{\rm GeV}
for ${\cal L}=50$ $ {\rm fb}^{-1}$.}
\end{center}
\end{figure}
For the sample analysis, we use two observables: $\sigma_{F}$ and
$\sigma_B$, the forward
and backward cross-sections.
The forward (backward) cross-section is defined by
the requirement that the angle between the momenta of the $e^-$ and
the $W^-$ is less (more) than $90^\circ$.
We choose these observables since they may be used to form the total
cross-section $\sigma_{tot} = \sigma_{F} + \sigma_B$ (arising from
cross-section parts even in the production angle) and the
forward backward
asymmetry 
$A_{FB} = ( \sigma_{F} - \sigma_B)/\sigma_{tot}$ 
(arising from odd cross-section parts).
For two different sets of anomalous couplings,
the two rings with allowed values derived from $\sigma_{F}$ and
$\sigma_B$ overlap almost
totally in the case of ${\cal P}$ conserving couplings $x_\gamma$,
$\delta_Z$.
When replacing $\delta_Z$ by the ${\cal P}$ violating coupling $z_Z$,
the allowed ranges overlap much less since the forward-backward
asymmetry is more sensitive to this coupling.
All this is nicely seen in figure~\ref{rings} (where we also took ISR
into account).
There, one further may notice that $\sigma_B$ is more
sensitive to anomalous couplings than $\sigma_F$ although the relative
statistical error of the latter is much smaller.
This is in accordance with the properties of the angular distributions
in figure~\ref{anoall}
(see also figure~2 in \cite{Biebel:1997id} for a center-of-mass energy of 
500 GeV).
A similar discussion has been performed for other pairs of anomalous
couplings in \cite{Biebel:1997ti}.

\section{Summary}
\ezero

We determined semi-analytical background and anomalous contributions
to the differential cross-section for $W$ pair
production in processes of the {\tt CC11} class.
With the $W$ production angle as an additional parameter, the expressions
are not as compact as those for the total cross-section.

By performing numerical calculations with the {\tt GENTLE} package we 
illustrated the effects of the background contributions.
At energies of about 500 GeV or more, background has sizeable effects
especially for backward scattering.
At energies of about 190~GeV background is less than 0.3\% for
all scattering angles.
Contributions from anomalous couplings are strongest in the same region.
Therefore, they cannot be studied without
taking background properly into account.

The present calculation has several limitations, of more importance at 
higher energies:
the virtual corrections are not taken into account; for angular
distributions, the QED ISR radiator function used is only an
approximation for the real photonic corrections (both with respect to
the $O(\alpha)$ part and to higher  
order corrections); the treatment of finite width effects may be
refined.
One should estimate the net effect to be of the order of up to few per
cent at $\sqrt{s} = 500$ GeV.
Thus, {\tt GENTLE} is certainly an appropriate tool for the study of 
$W$ pair production at LEP 2 while at higher energy it may serve as a
playing ground for sensitivity studies but may not replace a more
complete calculation.

\bigskip

\noindent
{\bf
Acknowledgement
}
\\
We would like to thank Th.~Ohl for numerous discussions,
hints, and numerical comparisons with the Fortran
program {\tt WOPPER} \cite{Anlauf:1994sc,Anlauf:1996wq}.
Further, we would like to thank D.~Bardin for the continuous fruitful
collaboration in the {\tt GENTLE} project.
\begin{appendix}
\def\theequation{\Alph{section}.\arabic{equation}}
\ezero
\section{The {\tt CC03} Process}
\label{CC03results}
\ezero
The coefficient functions used in (\ref{cc03diff}) are:
\begin{eqnarray}
  {\cal    C}^t&=&\frac{2}{(6\pi^2)^2}\mbox{Re}
  \frac{1}{|D_W(s_1)|^2|D_W(s_2)|^2}
\nonumber\\&&\mbox{}
  \times L^4(e,W))L^2(F_1,W)L^2(F_2,W)N_c(F_1)N_c(F_2),
\\
  {\cal C}^{st}&=&\sum\limits_{k=\gamma,Z}\frac{2}{(6\pi^2)^2}\mbox{Re}
  \frac{1}{|D_W(s_1)|^2|D_W(s_2)|^2D_{k}(s)}
\nonumber\\&&\mbox{}
  \times g_k L(e,l)L^2(F_1,W)L^2(F_2,W)L^2(e,W)N_c(F_1)N_c(F_2),
\\
  {\cal C}^{s}&=&\sum\limits_{k,l=\gamma,Z}\frac{2}{(6\pi^2)^2}\mbox{Re}
  \frac{1}{|D_W(s_1)|^2|D_W(s_2)|^2D_k(s)D^*_l(s)}
\nonumber\\&&\mbox{}
  \times g_kg_l[L(e,k)L(e,l)+R(e,k)R(e,l)]
\nonumber\\&&\mbox{}
  \times L^2(F_1,W)L^2(F_2,W)N_c(F_1)N_c(F_2).
\end{eqnarray}
The denominators of the boson propagators are:
\begin{equation}
D_V(s)=s-M_V^2+iM_V\Gamma_V,
\label{propag}
\end{equation}
and the coupling constants in the standard model are:
\begin{equation}
  \begin{array}{rclrcl}
\\
    g_\gamma&=&gs_W=e,&
    g_Z&=&gc_W,
\\
\\
    L(f,W)&=&\frac{\displaystyle g}{\displaystyle 2\sqrt{2}},
    \hspace{2cm}&
    R(f,W)&=&0,
\\
\\
    L(f,\gamma)&=&\frac{\displaystyle eQ_f}{\displaystyle 2},&
    L(f,Z)&=&\frac{\displaystyle e}{\displaystyle 4s_Wc_W}
    (2I_3^f-2Q_fs_W^2),
\\
\\
    R(f,\gamma)&=&\frac{\displaystyle eQ_f}{\displaystyle 2},&
    R(f,Z)&=&\frac{\displaystyle e}{\displaystyle 4s_Wc_W}(-2Q_fs_W^2).
\\
  \end{array}
\label{couplingconst}
\end{equation}
We use $Q_e=-1$ and $I^e_3=-\frac{1}{2}$.
The colour factor $N_c$ is~1 for leptons and~3 for quarks.

For the kinematical functions ${\cal G}$ we quote the expressions
from \cite{Bardin:1996uc}:
\begin{eqnarray}
  {\cal G}^t&=&\frac{1}{8}\left[2s(s_1+s_2)+\frac{\lambda}{4}\sin^2\theta
  +\frac{\lambda s_1s_2\sin^2\theta}{t^2_\nu}\right],
\\
\nonumber\\
  {\cal  G}^{st}&=&\frac{1}{8}\left[(s-s_1-s_2)\left(2s(s_1+s_2)+
  \frac{\lambda}{4}\sin^2\theta\right)\right.
\nonumber\\&&\mbox{}\left.
  -\frac{s_1s_2}{t_\nu}\left(4s(s_1+s_2)
  -\lambda\sin^2\theta\right)\right],
\\
\nonumber\\
  {\cal G}^s&=&\frac{1}{32}\lambda\left[2s(s_1+s_2)
  +\left(3s_1s_2+\frac{\lambda}{4}\right)\sin^2\theta\right],
\end{eqnarray}
with 
\begin{equation}
  \lambda=s^2+s_1^2+s_2^2-2ss_1-2ss_2-2s_1s_2
\end{equation}
and the denominator of the neutrino propagator $t_\nu$:
\begin{equation}
  t_\nu=\frac{1}{2}\left(s-s_1-s_2-\sqrt{\lambda}\cos\theta\right).
\label{nuprop}
\end{equation}

\section{QED Corrections}
\label{qedcorrections}
\ezero
The differential cross-sections are calculated in the
rest system $\Sigma'$ of the $W$ boson 
pairs (or, equivalently, of the final state fermion pairs). 
If energetic photons are radiated from the initial state, $\Sigma'$  
differs from the laboratory system  $\Sigma$ where 
the production angles are determined experimentally.
The corresponding Lorentz boost will be described in appendix \ref{lobo}. 
An emission of photons from $e^-$ or $e^+$ leads to different
relations between the $W$ production angle in $\Sigma$ and in
$\Sigma'$.
Thus, we have to use the structure function approach for a description
of ISR since here the energy loss of each initial state particle is known.
\subsection{Structure function approach}
\label{SFA}
In the structure function approach \cite{Kuraev:1985hb,Beenakker:1996kt},
the initial state photonic corrections
are taken into account by convoluting the tree-level cross-section twice
with the structure function $D(x,s)$ (with the structure functions
as described in section 2.3 of \cite{Bardin:1996zz} and references therein):
\begin{eqnarray}
  \frac{\mbox{d}\sigma_{\mbox{\scriptsize QED}}(s)}
  {\mbox{d}s_1\mbox{d}s_2\mbox{d}\cos\theta}
  &=&\int\limits_{x_1^{\mbox{\tiny min}}}^1\!\!\!\mbox{d}x_1
  \int\limits_{x_2^{\mbox{\tiny min}}}^1\!\!\!\mbox{d}x_2
  D(x_1,s)D(x_2,s)
\sum_{i=1,2}\left|\frac{\mbox{d}\cos\theta'_{i}}
  {\mbox{d}\cos\theta}\right|
  \frac{\mbox{d}\sigma(x_1x_2s,s_1,s_2)}{\mbox{d}\cos\theta'_{i}}
  ,
\nonumber\\
\label{qeddiff}
\end{eqnarray}
with $\theta'_i=\theta'_i(s,s_1,s_2,x_1,x_2,\theta)$ and the
lower integration boundaries
\begin{eqnarray}
  x_1^{\mbox{\tiny min}}&\geq&\frac{(\sqrt{s_1}+\sqrt{s_2})^2}{s},
\\
  x_2^{\mbox{\tiny min}}&\geq&\frac{(\sqrt{s_1}+\sqrt{s_2})^2}{x_1s}.
\end{eqnarray}
The sum in (\ref{qeddiff}) indicates that no, one, or two
solutions may exist for $\theta'_i$ (defined in $\Sigma '$) at given
values of the parameters in $\Sigma$.
The Jacobean is easily derived from (\ref{jaco}):
\begin{eqnarray}
  \frac{{\rm d}\cos\theta'_{1,2}}{{\rm
      d}\cos\theta}&=&\frac{\beta_{1,2}(1-v^2)} 
  {\left[\beta_{1,2}^2+v^2(1-\beta_{1,2}^2\sin^2\theta)-2v\beta_{1,2}
      \cos\theta\right]^{3/2}}\\
  &&\times\left[\beta_{1,2}-v\cos\theta\pm  v(1-\cos^2\theta)\frac{1-b^2}
     {b}\frac{v}{1\pm vb\cos\theta}\right].
\label{jacob}
\end{eqnarray}
The $b$ is given in (\ref{bdef}), the velocity $v$ 
of the $W^+W^-$ system  in the laboratory frame $\Sigma$ in (\ref{vdef}), 
and the velocities $\beta_{1,2}$  
of the $W^-$  in $\Sigma$ in 
   (\ref{beta-i}).
For doubly resonant diagrams the cross-section has to be multiplied by the 
Coulomb correction $C(x_1x_2s)$
\cite{Fadin:1993kg,Bardin:1993mc,Fadin:1995pm}; 
we follow \cite{Bardin:1993mc} as described in \cite{Bardin:1996uc}. 

For applications and comparisons with Monte Carlo programs
\cite{Bardin:1997gc,Ohl:1996ig}
it might be more convenient to determine not the 
differential cross-section itself but to perform a bin-wise integration:
\begin{eqnarray}
\sigma = \sum_i
\int_{\cos\theta_{a_i}'(\theta_a)}^{\cos\theta_{b_i}'(\theta_b)}  
\frac{d \sigma}{d\cos\theta'}.
\end{eqnarray}
Such an integration may be trivially performed analytically in $\Sigma'$ 
in view of the 
relatively simple angular dependencies and computer time may be saved.
Of course, the boosted integration boundaries have to be determined.
For a given angular bin in the laboratory system, there may exist zero, one, 
or two bins to be integrated over in the boosted frame.  
More details on this may be found in section 2.4 of \cite{Bardin:1996zz}.
The bin-integrated cross-sections are used in {\tt GENTLE} for {\tt
CC03} processes.

Finally, a remark on the use of the structure function $D(x,s)$ might be necessary.
This structure function is determined for the total cross-section
only.
Thus, for the differential cross-section it has to be considered as an 
approximation.
\subsection{Lorentz boost}
\label{lobo}
We will denote 4-momenta in $\Sigma'$
as $p'$ and in $\Sigma$ 
as $p$.
In $\Sigma$, the momenta of electron and positron are:
\begin{eqnarray}
  p_{e^-}&=&Ex_1(1,0,0,1),\label{e-d}\\
  p_{e^+}&=&Ex_2(1,0,0,-1).\label{e+d}
\end{eqnarray}
$E=\sqrt{s}/2$ denotes the beam energy.
In  $\Sigma'$, the sum of the spatial momenta of the two particles
vanishes and one gets in this frame:
\begin{eqnarray}
  p'_{e^-}&=&E\sqrt{x_1x_2}(1,0,0,1),\\
  p'_{e^+}&=&E\sqrt{x_1x_2}(1,0,0,-1).
\end{eqnarray}
Applying the transformation formula
\begin{equation}
  p'_3=\frac{p_3-vp_0}{\sqrt{1-v^2}}\label{lotra}
\end{equation}
on one of the beam particles, one may derive the relative velocity of the two
Lorentz frames
\begin{equation}
  v=\frac{x_1-x_2}{x_1+x_2}.
\label{vdef}
\end{equation}
In  $\Sigma'$, one may choose the momenta of the $W$ bosons as follows:
\begin{eqnarray} 
  p'_{W^-}&=&\left(\sqrt{\frac{\lambda'}{4s'}+s_1},\,\,
  \sqrt{\frac{\lambda'}{4s'}}\sin\theta',\,\,0,\,\,
  \sqrt{\frac{\lambda'}{4s'}}\cos\theta'\right),
\label{ppmin}
\\
  p'_{W^+}&=&\left(\sqrt{\frac{\lambda'}{4s'}+s_2},\,\,
 - \sqrt{\frac{\lambda'}{4s'}}\sin\theta',\,\,0,\,\,
 - \sqrt{\frac{\lambda'}{4s'}}\cos\theta'\right),
\end{eqnarray}
where $s'=4x_1x_2E^2$ is the reduced center-of-mass energy and
\begin{equation}
  \lambda'\equiv\lambda(s',s_1,s_2).
\end{equation}

The energy and momenta of the bosons are fixed for given values of
$s_1$ and $s_2$.
The momentum of the $W^-$ boson in  $\Sigma$ can
be written as:
\begin{equation}
  p_{W^{-}}=(Q_i,B_i\sin\theta,0,B_i\cos\theta)\label{momW},
\end{equation}
where $Q_i$ and $B_i$ are real and positive functions of $s'$,
$s_1$, $s_2$, $\cos\theta$ and $v$.
The velocity of the $W^{-}$-boson in the laboratory system is then:
\begin{eqnarray}
  \beta_{i}=\frac{B_{i}}{Q_{i}}\, ,\hspace{2cm} i=1,2.
\label{beta-i}
\end{eqnarray}
With 
\begin{equation}
p_{e^-}+p_{e^+}=p_{W^-}+ p_{W^+}
\end{equation}
and 
\begin{equation}
p_{W^-}^2=s_1,\mbox{\hspace{3cm}}p_{W^+}^2=s_2,
\end{equation}
two sets of solutions may be obtained:
\begin{eqnarray}
  B_{1,2}&=&\frac{(s'-s_2+s_1)\sqrt{1-v^2}\left(v\cos\theta
    \pm b\right)}
    {2\sqrt{s'}(1-v^2\cos^2\theta)},\\
  Q_{1,2}&=&\frac{(s'-s_2+s_1)\sqrt{1-v^2}\left(1\pm bv\cos\theta
    \right)}
      {2\sqrt{s'}(1-v^2\cos^2\theta)}.
\end{eqnarray}
Here, we used the abbreviation 
\begin{equation}
    b=\sqrt{1-\frac{4s_1s'(1-v^2\cos^2\theta)}{(s'-s_2+s_1)^2(1-v^2)}}.
\label{bdef}
\end{equation}

The number of solutions depends on $\cos\theta$, $v$ and $b$.
By definition, $B$ is real and positive.
There is no solution, when $v\cos\theta<-b$, one solution for
$|v\cos\theta|<b$, and two solutions exist for $v\cos\theta>b$.

With the given solutions for $B$ and $Q$ and eq.~(\ref{lotra}),
the relation between the $W$ production angles in the
two Lorentz systems is found:
\begin{equation}
  \cos\theta'=\frac{B\cos\theta-vQ}{\sqrt{(1-v^2)B^2\sin^2\theta+
      (B\cos\theta-vQ)^2}}.
\label{jaco}
\end{equation}
For the limiting case of on-shell $W$ pair production the
transformation (\ref{jaco}) is
in accordance with a similar transformation given in
\cite{Beenakker:1989km}.  

\section{Neutral Current Kinematical Functions}
\label{NC20}
\ezero

In section \ref{purebackground}, we use two kinematical functions
known from the the study of neutral current process
\cite{Leike:1995kj}: 

The function ${\cal G}^{DD}_{422}$ is:
\begin{eqnarray}
  {\cal G}^{DD}_{422}(\cos\theta;s_1;s_2,s)&=&\frac{3}{8}(1+\cos^2\theta)
  {\cal G}_{422}(s_1;s_2,s)
\\&&\mbox{}
  +\frac{1-3\cos^2\theta}{\lambda}s_1(s+s_2)\frac{3}{4}\left(
  1-2{\cal L}(s_1;s_2,s)\frac{ss_2}{s_1-s_2-s}\right)\nonumber,
\end{eqnarray}
where ${\cal G}_{422}$ is also known from different contexts~
(\cite{Bardin:1994sc},%
\cite{Baier:1966}-
\nocite{Cvetic:1992qv,Stuart:1996sh,Stuart:1995ba,Stuart:1997nu}%
\cite{Stuart:1995zr}):\footnote{
We explicitely agree with \cite{Bardin:1994sc,Baier:1966,Cvetic:1992qv}.
}
\begin{equation}
  {\cal G}_{422}(s;s_1,s_2)=\frac{s^2+(s_1+s_2)^2}{s-s_1-s_2}
  {\cal L}(s;s_1,s_2)-2.
\label{g422}
\end{equation}

The function ${\cal G}^{DD}_{233}(\cos\theta,s,s_1,s_2)$ is:
\begin{eqnarray}
  {\cal G}_{233}^{DD}(\cos\theta,s,s_1,s_2)&\!=\!&\frac{3}{8}(1+\cos^2\theta)
  {\cal G}_{233}(s;s_1,s_2)
\nonumber\\&&\mbox{}
  -\frac{3}{\lambda^2}\frac{3}{8}(1-3\cos^2\theta)s\left[
  {\cal L}(s_1;s_2,s)2s_2(s_1-s_2)+(s-s_1-3s_2)\right]
\nonumber\\&&\mbox{}\vphantom{\frac{3}{8}}
  \times\left[{\cal L}(s_2;s,s_1)2s_1(s_2-s_1)+(s-s_2-3s_1)\right],
\end{eqnarray}
with \cite{Bardin:1995vm}
\begin{eqnarray}
  {\cal G}_{233}(s;s_1,s_2)&=&
  \frac{3}{\lambda^2}\left\{{\cal L}(s_2;s,s_1){\cal L}(s_1;s_2,s)\right.
\nonumber\\&&\mbox{}
  4s\left[ss_1(s-s_1)^2+ss_2(s-s_2)^2+s_1s_2(s_1-s_2)^2\right]
\nonumber\\&&\mbox{}
  +(s+s_1+s_2)\left[{\cal
    L}(s_2;s,s_1)2s\left[(s-s_2)^2+s_1(s+s_2-2s_1)\right]\right.
\nonumber\\&&\mbox{\hspace{2.5cm}}
  +{\cal L}(s_1;s_2,s)2s\left[(s-s_1)^2+s_2(s+s_1-2s_2)\right]
\nonumber\\&&\mbox{\hspace{2.5cm}}\left.\left.
  +5s^2-4s(s_1+s_2)-(s_1-s_2)^2\right]\right\}.
\end{eqnarray}

\end{appendix}

\end{fmffile}

\newpage
\small


\begingroup\begin{thebibliography}{10}

\bibitem{Glashow:1961ez}
S.~L. Glashow, {\em Nucl. Phys.} {\bf 22} (1961) 579.

\bibitem{Weinberg:1967pk}
S.~Weinberg, {\em Phys. Rev. Lett.} {\bf 19} (1967) 1264.

\bibitem{Salam:1968rm}
A.~Salam, ``{Weak and Electromagnetic Interactions}'', in {\em Proc. of the
  Nobel Symposium, 1968, Lerum, Sweden} (N.~Svartholm, ed.), pp.~367--377,
  Almqvist and Wiksell, Stockholm, 1968.

\bibitem{Flambaum:1975wp}
V.~Flambaum, I.~Khriplovich, and O.~Sushkov, {\em Sov. J. Nucl. Phys.} {\bf 20}
  (1975) 537--540.

\bibitem{Alles:1977qv}
W.~Alles, C.~Boyer, and A.~J. Buras, {\em Nucl. Phys.} {\bf B119} (1977) 125.

\bibitem{Tsai:1965hq}
Y.-S. Tsai and A.~C. Hearn, {\em Phys. Rev.} {\bf 140} (1965) B721--B729.

\bibitem{Muta:1986is}
T.~Muta, R.~Najima, and S.~Wakaizumi, {\em Mod. Phys. Lett.} {\bf A1} (1986)
  203.

\bibitem{Berends:1994pv}
F.~A. Berends, R.~Pittau, and R.~Kleiss, {\em Nucl. Phys.} {\bf B424} (1994)
  308--342.

\bibitem{Bardin:1994sc}
D.~Bardin, M.~Bilenky, D.~Lehner, A.~Olchevski, and T.~Riemann, {\em Nucl.
  Phys. (Proc. Suppl.) 37B} (1994) 148--157.

\bibitem{Beenakker:1996kt}
W.~Beenakker {\em et~al.}, ``{$WW$} cross-sections and distributions'', in {\em
  Physics at {LEP2}, {\rm CERN 96--01 (1996)}} (G.~Altarelli, T.~{Sj\"ostrand},
  and {F. Zwirner}, eds.), pp.~79--139.

\bibitem{Bardin:1986fi}
D.~Bardin, S.~Riemann, and T.~Riemann, {\em Z. Phys.} {\bf C32} (1986) 121.

\bibitem{Jegerlehner:1986vs}
F.~Jegerlehner, {\em Z. Phys.} {\bf C32} (1986) 425; E: ibid., {\bf C38} (1988)
  519.

\bibitem{Bohm:1988ck}
W.~Beenakker, F.~Berends, M.~{B\"ohm}, A.~Denner, H.~Kuijf, and T.~Sack, {\em
  Nucl. Phys.} {\bf B304} (1988) 463.

\bibitem{Fleischer:1989kj}
J.~Fleischer, F.~Jegerlehner, and M.~Zralek, {\em Z. Phys.} {\bf C42} (1989)
  409.

\bibitem{Denner:1990tx}
A.~Denner and T.~Sack, {\em Z. Phys.} {\bf C46} (1990) 653.

\bibitem{Beenakker:1991sf}
W.~Beenakker, K.~Kolodziej, and T.~Sack, {\em Phys. Lett.} {\bf B258} (1991)
  469--474.

\bibitem{Dittmaier:1992np}
S.~Dittmaier, M.~B{\"o}hm, and A.~Denner, {\em Nucl. Phys.} {\bf B376} (1992)
  29--51.

\bibitem{Fleischer:1993nw}
J.~Fleischer, K.~Kolodziej, and F.~Jegerlehner, {\em Phys. Rev.} {\bf D47}
  (1993) 830--836.

\bibitem{Beenakker:1997bp}
W.~Beenakker, A.~P. Chapovsky, and F.~A. Berends, {\em Phys. Lett.} {\bf B411}
  (1997) 203.

\bibitem{Beenakker:1997ir}
W.~Beenakker, A.~P. Chapovsky, and F.~A. Berends, {\em Nucl. Phys.} {\bf B508}
  (1997) 17.

\bibitem{Denner:1997ia}
A.~Denner, S.~Dittmaier, and M.~Roth, {\em Nucl. Phys.} {\bf B519} (1998) 39.

\bibitem{Gaemers:1979hg}
K.~J.~F. Gaemers and G.~J. Gounaris, {\em Z. Phys.} {\bf C1} (1979) 259.

\bibitem{Hagiwara:1987vm}
K.~Hagiwara, R.~D. Peccei, D.~Zeppenfeld, and K.~Hikasa, {\em Nucl. Phys.} {\bf
  B282} (1987) 253.

\bibitem{Hagiwara:1993ck}
K.~Hagiwara, S.~Ishihara, R.~Szalapski, and D.~Zeppenfeld, {\em Phys. Rev. D}
  {\bf 48} (1993) 2182--2203.

\bibitem{Bilchak:1984ur}
C.~L. Bilchak and J.~D. Stroughair, {\em Phys. Rev.} {\bf D30} (1984) 1881.

\bibitem{Jegerlehner:1994zp}
F.~Jegerlehner, {\em Nucl. Phys. (Proc. Suppl.) 37B} (1994) 129--140.

\bibitem{HarunarRashid:1994mp}
A.~M.~H. ar~Rashid and K.~S. Islam, {\em Int. J. Mod. Phys.} {\bf A9} (1994)
  2783--2804.

\bibitem{Gounaris:1996rz}
G.~Gounaris {\em et~al.}, ``Triple gauge boson couplings'', in {\em Physics at
  {LEP2}, {\rm CERN 96--01 (1996)}} (G.~Altarelli, T.~{Sj\"ostrand}, and {F.
  Zwirner}, eds.), pp.~525--576.

\bibitem{Accomando:1997wt}
{ECFA/DESY LC Physics Working Group} Collaboration, E.~Accomando {\em et~al.},
  {\em Phys. Rept.} {\bf 299} (1998) 1.

\bibitem{Clare:1998aa}
R.~Clare, {\em Acta Phys. Polon.} {\bf B29} (1998) 2667.

\bibitem{Ellison:1998uy}
J.~Ellison and J.~Wudka, ``Study of trilinear gauge boson couplings at the
  {T}evatron collider'', UC Riverside preprint UCR-D0-98-01 (1998), subm. to
  Ann. Rev. Nucl. Part. Sci. \href{http://xxx.lanl.gov/abs/e-print
  hep-ph/9804322}{{\tt e-print hep-ph/9804322}}.

\bibitem{Ellison:1998ub}
{CDF and D0} Collaborations, J.~Ellison, ``Measurements of the {W} boson mass
  and trilinear gauge boson couplings at the {T}evatron'', UC Riverside
  preprint UCR-D0-98-21, June 1998, to be published in Proc. XXXIII$^d$
  Rencontres de Moriond, Electroweak Interactions and Unified Theories, Les
  Arcs, Savoie, France, March 14-21 1998 \href{http://xxx.lanl.gov/abs/e-print
  hep-ex/9806004}{{\tt e-print hep-ex/9806004}}.

\bibitem{Gounaris:1997}
G.~Gounaris and C.~Papadopoulos, ``Studying Trilinear Gauge Couplings in
  $e^+e^-\to l^-\bar{\nu}_lq\bar{q}'$ at Linear Collider Energies'', in
  {\em $e^+ e^-$ Linear Colliders: Physics and Detector
  Studies - Part E, {\rm DESY 97-123E (1997)}} (R.~Settles, ed.), p.~143.

\bibitem{Bardin:1993nb}
D.~Bardin, A.~Olshevsky, M.~Bilenky, and T.~Riemann, {\em Phys. Lett.} {\bf
  B308} (1993) 403--410; E: ibid., {\bf B357} (1995) 725.

\bibitem{Bardin:1996jw}
D.~Bardin, D.~Lehner, and T.~Riemann, {\em Nucl. Phys.} {\bf B477} (1996)
  27--58.

\bibitem{Bardin:1996uc}
D.~Bardin and T.~Riemann, {\em Nucl. Phys.} {\bf B462} (1996) 3--28.

\bibitem{Bardin:1995vm}
D.~Bardin, A.~Leike, and T.~Riemann, {\em Phys. Lett.} {\bf B344} (1995)
  383--390.

\bibitem{Bardin:1996zz}
D.~Bardin, J.~Biebel, D.~Lehner, A.~Leike, A.~Olchevski, and T.~Riemann, {\em
  Comput. Phys. Commun.} {\bf 104} (1997) 161. \\{{\tt GENTLE} is available at
  {\tt http://www.ifh.de/theory/publist.html}}.

\bibitem{Vermaseren:1991}
J.~A.~M. Vermaseren, {``Symbolic Manipulation with {\tt FORM}}'' (Computer
  Algebra Nederland, Amsterdam, 1991).

\bibitem{Argyres:1995ym}
E.~N. Argyres {\em et~al.}, {\em Phys. Lett.} {\bf B358} (1995) 339--346.

\bibitem{Beenakker:1996kn}
W.~Beenakker {\em et~al.}, {\em Nucl. Phys.} {\bf B500} (1997) 255.

\bibitem{Biebel:1997id}
J.~Biebel, ``{Four fermion production with anomalous couplings at LEP 2 and
  NLC}'', in {\em {Proc. of XIIth International Workshop on High Energy Physics
  and Quantum Field Theory, 4--10 Sep~1997, Samara, Russia}} (B.~Levtchenko,
  ed.), 1998, to appear; e-print hep-ph/9711439.

\bibitem{Berends:1995dn}
F.~A. Berends and A.~I. van Sighem, {\em Nucl. Phys.} {\bf B454} (1995)
  467--484.

\bibitem{Aronson:1969aa}
H.~Aronson, {\em Phys. Rev.} {\bf 186} (1969) 1434--1441.

\bibitem{Anlauf:1994sc}
H.~Anlauf, J.~Biebel, H.~Dahmen, A.~Himmler, P.~Manakos, T.~Mannel, and
  W.~Sch{\"o}nau, {\em Comput. Phys. Commun.} {\bf 79} (1994) 487--502.

\bibitem{Anlauf:1996wq}
H.~Anlauf, P.~Manakos, T.~Ohl, and H.~Dahmen, ``{{\tt WOPPER}, version 1.5: A
  Monte Carlo event generator for $e^+e^- \rightarrow (W^+ W^-) \rightarrow 4 f
  + n \gamma$ at LEP-2 and beyond}'', Darmstadt preprint IKDA 96--15 (1996),
  \href{http://xxx.lanl.gov/abs/e-print hep-ph/9605457}{{\tt e-print
  hep-ph/9605457}}.

\bibitem{Biebel:1997ti}
J.~Biebel and T.~Riemann, ``Semianalytic predictions for {$W$} pair production
  at 500 {GeV}'', in {\em $e^+ e^-$ Linear Colliders: Physics and Detector
  Studies - Part E, {\rm DESY 97-123E (1997)}} (R.~Settles, ed.), p.~139.

\bibitem{Kuraev:1985hb}
E.~A. Kuraev and V.~S. Fadin, {\em Sov. J. Nucl. Phys.} {\bf 41} (1985)
  466--472.

\bibitem{Fadin:1993kg}
V.~S. Fadin, V.~A. Khoze, and A.~D. Martin, {\em Phys. Lett.} {\bf B311} (1993)
  311--316.

\bibitem{Bardin:1993mc}
D.~Bardin, W.~Beenakker, and A.~Denner, {\em Phys. Lett.} {\bf B317} (1993)
  213--217.

\bibitem{Fadin:1995pm}
V.~S. Fadin, V.~A. Khoze, A.~D. Martin, and A.~Chapovsky, {\em Phys. Rev.} {\bf
  D52} (1995) 1377--1385.

\bibitem{Bardin:1997gc}
D.~Bardin {\em et~al.}, ``Event generators for {$WW$} physics'', in {\em
  Physics at {LEP2}, {\rm CERN 96--01 (1996)}} (G.~Altarelli, T.~{Sj\"ostrand},
  and {F. Zwirner}, eds.), vol.~2, pp.~3--353.

\bibitem{Ohl:1996ig}
T.~Ohl, {\em Acta Phys. Polon.} {\bf B28} (1997) 847.

\bibitem{Beenakker:1989km}
W.~Beenakker, F.~A. Berends, and W.~L. van Neerven, ``Applications of
  renormalization group methods to radiative corrections'', in {\em {Proc. of
  Workshop on Electroweak Radiative Corrections for e+ e- Collisions, 3--7
  April 1989, Tegernsee, Germany}} ({J.~H.~K\"uhn}, ed.), pp.~3--24,
  Springer-Verlag, Berlin, 1989.

\bibitem{Leike:1995kj}
A.~Leike, ``{Semianalytic distributions in four fermion neutral current
  processes}'', in {\em {Proc. of Int. Workshop on Perspectives for Electroweak
  Interactions in $e^+e^-$ Collisions, 5--8 Feb. 1995, Tegernsee, Germany}}
  (B.~Kniehl, ed.), pp.~121--130, World Scientific, Singapore, 1995.

\bibitem{Baier:1966}
V.~Ba\u{\i}er, V.~Fadin, and V.~Khoze, {\em Sov. J. Nucl. Phys.} {\bf 23}
  (1966) 104--111.

\bibitem{Cvetic:1992qv}
M.~{Cveti\v{c}} and P.~Langacker, {\em Phys. Rev.} {\bf D46} (1992) 4943--4954.

\bibitem{Stuart:1996sh}
R.~Stuart, ``Gauge invariance in boson production'', in {\em {Proc. of Int.
  Workshop on the Higgs Puzzle, 8--13 Dec. 1996, Tegernsee, Germany}}
  (B.~Kniehl, ed.), pp.~47--54, World Scientific, Singapore, 1996.

\bibitem{Stuart:1995ba}
R.~Stuart, ``Unstable particles'', in {\em {Proc. of Int. Workshop on
  Perspectives for Electroweak Interactions in $e^+e^-$ Collisions, 5--8 Feb.
  1995, Tegernsee, Germany}} (B.~Kniehl, ed.), pp.~235--246, World Scientific,
  Singapore, 1995.

\bibitem{Stuart:1997nu}
R.~Stuart, ``Gauge invariance and the unstable particle'', talks at {\it
  Workshop on Hadron Production Cross-Sections at DAPHNE meeting, 1--2 Nov
  1996, Karlsruhe, Germany}, and at {\it 1st Latin American Symposium on
  High-energy Physics, {\rm SILAFAE-I}, 1--5 Nov 1996, Merida, Mexico},
  unpublished, {\tt hep-ph/9706550}.

\bibitem{Stuart:1995zr}
R.~Stuart, {\em Nucl. Phys.} {\bf B498} (1997) 28.

\end{thebibliography}\endgroup
\end{document}